\newif\ifreviewstyle
\reviewstylefalse

\ifreviewstyle
  \documentclass[onecolumn]{IEEEtran}
\else
  \documentclass[journal]{IEEEtran}
\fi

\usepackage{cite}
\usepackage[pdftex]{graphicx}
\usepackage{mathtools}
\usepackage{amsmath,amssymb}
\usepackage{amsthm} % <-- gives \newtheorem and the proof environment
\usepackage{bbold}
\usepackage{array}
\usepackage{url}
\usepackage{enumitem}
\usepackage[normalem]{ulem}
\usepackage[capitalize,noabbrev,nameinlink]{cleveref}

\setlist[enumerate,1]{label=\textit{\arabic*)}, ref=\textit{\arabic*)}}
\setlist[enumerate,2]{label=\alph*), ref=\textit{\arabic{enumi})}~\alph*)}

\theoremstyle{plain}
\newtheorem{theorem}{Theorem}

\newtheorem{lemma}[theorem]{Lemma}
\newtheorem{corollary}[theorem]{Corollary}

\theoremstyle{definition}
\newtheorem{definition}[theorem]{Definition}
\newtheorem{assumption}[theorem]{Assumption}
\newtheorem{notation}[theorem]{Notation}
\newtheorem{remark}[theorem]{Remark}

\newcommand{\px}{p_X}
\newcommand{\pyx}{p_{N}}
\newcommand{\py}{p_Y}

\newcommand{\pa}[1]{\left(#1\right)}      % parenthesis
\newcommand{\pasq}[1]{\left[#1\right]}   % squared parenthesis
\newcommand{\pacu}[1]{\left\{#1\right\}} % curly parenthesis

\newcommand{\dhat}[1]{\hat{d}_{#1}}
\newcommand{\dbar}[1]{\bar{d}_{#1}}
\newcommand{\mhat}[1]{\hat{m}_{#1}}
\newcommand{\mbar}[1]{\bar{m}_{#1}}
\newcommand{\zhat}{\hat{z}}
\newcommand{\zbar}{\bar{z}}
\newcommand{\cbar}{\bar{c}}
\newcommand{\Mkg}[1]{M_{#1}^{>}}
\newcommand{\Mkl}[1]{M_{#1}^{<}}
\newcommand{\mkg}[1]{m_{#1}^{>}}
\newcommand{\mkl}[1]{m_{#1}^{<}}
\newcommand{\xkg}[1]{S_{#1}^{>}}
\newcommand{\xkl}[1]{S_{#1}^{<}}
\newcommand{\Ex}[1]{\big\langle#1\big\rangle}
\newcommand{\func}[1]{{\cal L}_{#1}}

\newcommand{\range}[2]{[#1\!:\!#2]}   % [k:l]
\newcommand{\rangeone}[1]{[#1]}         % [l] = [1:l]

\DeclarePairedDelimiter\floor{\lfloor}{\rfloor}

\newenvironment{casenv}
{\begin{enumerate}[label=\Roman*., ref=\Roman*]}
{\end{enumerate}}

\AtBeginDocument{\providecommand\figref[1]{Fig.~\ref{fig:#1}}}
\AtBeginDocument{\providecommand\corref[1]{Corollary \ref{cor:#1}}}
\AtBeginDocument{\providecommand\lemref[1]{Lemma \ref{lem:#1}}}
\AtBeginDocument{\providecommand\thmref[1]{Theorem \ref{thm:#1}}}
\AtBeginDocument{}
\AtBeginDocument{}

\newlength{\figW}
\ifreviewstyle
\else
\fi

\begin{document}
\title{Phase Transitions of the Additive Uniform Noise Channel with Peak Amplitude and Cost Constraint}

\author{Jonas~Stapmanns,
        Luke~Eilers,
        Catarina~Dias,
        Tobias~K\"uhn,
        and~Jean-Pascal~Pfister% <-this % stops a space
\thanks{This work was presented in part at the IEEE International Symposium on Information Theory (ISIT)~2025.}
\thanks{All authors are with the Department of Physiology, University of Bern, Bern, Switzerland (e-mail: \{jonas.stapmanns, catarina.reisdias, luke.eilers, tobias.kuehn, jeanpascal.pfister\}@unibe.ch.}% <-this % stops a space
\thanks{}}

\markboth{}%
{Stapmanns \MakeLowercase{\textit{et al.}}: Capacity-Achieving Input Distribution of the Additive Uniform Noise Channel With Peak Amplitude and Cost Constraint}

\maketitle

\begin{abstract}
Under which condition is quantization optimal?
We address this question in the context of the additive uniform noise channel under peak amplitude and cost constraints.
We compute analytically the capacity-achieving input distribution as a function of the noise level, the average cost constraint, and the curvature of the cost function.
We find that when the cost function is concave, the capacity-achieving input distribution is discrete, whereas when the cost function is convex and the cost constraint is active, the support of the capacity-achieving input distribution spans the entire interval.
For the cases of a discrete capacity-achieving input distribution, we derive the analytical expressions for the capacity of the channel.
\end{abstract}


\IEEEpeerreviewmaketitle

\section{Introduction}

\IEEEPARstart{S}{ince} Shannon introduced channel capacity\cite{shannon_mathematical_1948}, capacity-achieving input distributions have been studied for several combinations of channels and constraints \cite{shannon_mathematical_1948,smith_information_1971,oettli_capacity-achieving_1974,shamai_capacity_1995,lapidoth_capacity_2009,dytso_when_2018,dytso_capacity_2019,dytso_capacity_2020,eisen_capacity-achieving_2023,barletta_binomial_2024}. In many cases, especially under a peak amplitude (PA) constraint, the capacity-achieving input distribution is discrete and consists of finitely many mass points \cite{fahs_2018}. Examples include the additive Gaussian channel \cite{smith_information_1971}, the Poisson channel \cite{lapidoth_capacity_2009,dytso_poisson_2021}, and the additive channel with piecewise constant noise \cite{oettli_capacity-achieving_1974}. Known exceptions are a few channels without PA such as the additive Gaussian channel with variance constraint \cite{shannon_mathematical_1948}, the Cauchy channel under logarithmic constraint \cite{fahs_2014}, and the exponential channel with non-negative input \cite{anantharam_1996}, which have a capacity-achieving distribution with continuous support.
%Non-negative PAs are analyzed less commonly due to their asymmetry but are relevant for optical intensity channels and biological systems, where signals must be non-negative due to physical limitations \cite{chan_capacity-achieving_2005, farsad_2020}.

What conditions determine whether the capacity-achieving input distribution is discrete? Is there a channel that exhibits a phase transition between a continuous and a discrete capacity-achieving input distribution? The first question has been studied by Das \cite{das_2000}, Tchamkerten \cite{tchamkerten_discreteness_2004}, and Fahs and Abou-Faycal \cite{fahs_2018}. However, their non-constructive approaches are not readily amenable to a detailed analysis of possible phase transitions. 
%Because of its analytical tractability, we frame those questions in the context of the additive uniform noise channel with non-negative PA and cost constraint.
To be able to investigate this open question analytically, we here focus on the additive uniform noise channel with non-negative PA and cost constraint.
We find that when the cost function is concave, the capacity-achieving input distribution is discrete, whereas it has continuous support when the cost function is convex and the cost constraint is active.

\section{Problem statement}
We investigate the capacity-achieving input distribution of the additive uniform noise channel 
\begin{equation}
Y=X+N,\quad\mathrm{where}\,N\sim\mathrm{Uniform}\left(-b,b\right),\label{eq:cahnnel_definition}
\end{equation}
with $b>0$. Hence, the density of the noise is given by $p_{N}\pa{z}=\mathbb{1}_{-b<z<b}/\pa{2b}$.
For convenience, we define an additional variable for the inverse width $r\coloneqq1/\left(2b\right)$.
We denote by $\px$ and $\py$ the probability distributions corresponding to $X$ and $Y$, respectively.
To emphasize that the output distribution is induced by $\px$, we write $\py\pa{y;\px}$ when needed.
The input to the channel is subject to the PA constraint $P\left(X<0\right)=P\left(X>1\right)=0$ and, additionally, to the cost constraint
\begin{equation}
    \Ex{c_{\alpha}(x)}\leq\bar{c}.\label{eq:cost_constraint}
\end{equation}
Unless specified otherwise, the expectation $\Ex{\cdot}$ is with respect to the input distribution that will be denoted as $\px$.
The cost function $c_\alpha\pa{x}$ assigns costs to the input of the channel and is specified below.
\begin{assumption}[Cost function]\label{ass:cost_function}
Let $\alpha>0$. The function $c_\alpha:\pasq{0,1}\rightarrow\pasq{0,1}$ is differentiable, strictly increasing, and
\begin{enumerate}
  \item strictly concave for $0<\alpha<1$,
  \item linear for $\alpha=1$, and
  \item strictly convex on a finite interval $G\subseteq\pasq{0,1}$ for $\alpha>1$.
\end{enumerate}
\end{assumption}
Thus, the cost function penalizes inputs near $x=1$ more heavily and feasible solutions for the input distribution have to reduce the probability mass of such high-cost inputs as the budget $\cbar$ tightens.
Without loss of generality we assume $c_{\alpha}\pa{0}=0$ and $c_{\alpha}\pa{1}=1$.
Henceforth, we will drop the index $\alpha$ and simply write $c\pa{x}$.
Throughout the paper we will use
\begin{equation}
    c\pa{x}=x^{\alpha}
\end{equation}
as an example in the figures, but all proofs are valid for the general cost function as defined in \Cref{ass:cost_function}.
The goal we pursue in this paper is to find the input distribution $\px^{\ast}$ that, for a given distribution of $Y$ conditional on $X$, i.e. $\pyx$, maximizes the mutual information between $X$ and $Y$, while satisfying the PA constraint and the cost constraint $\eqref{eq:cost_constraint}$. Enforcing this condition and the normalization of the probability distribution by Lagrange multipliers, this yields the formal definition of our question as an optimization problem:
\begin{definition}
    \label{Def_Ls}
    We define the functional
    \begin{equation}
        \func{}\pasq{\px,\nu,\lambda} \coloneqq \func{0}\pasq{\px,\nu} - \lambda \pa{\int_0^1 dx\, \px\pa{x}c\pa{x}-\cbar},\label{eq:def_constr_func}
    \end{equation}
    where
    \begin{equation}
    \begin{split}
        \func{0}\pasq{\px,\nu} \coloneqq& \int dy \,\int dx \, \pyx\pa{y-x}\,\px\pa{x}\log\pa{\frac{\pyx\pa{y-x}}{\py\pa{y}}}\\
        &\;- \nu \pa{\int_{0}^{1} dx\,\px\pa{x}-1}
        \end{split}\label{eq:def_unconstr_func}
    \end{equation}
    contains the mutual information between $X$ and $Y$ in the first term and the normalization condition for the input distribution $\px$ with the Lagrange multiplier $\nu$ in the second term.
    The second term in \eqref{eq:def_constr_func} represents the cost constraint through the Lagrange multiplier $\lambda$.
    With these definitions, we can express the objective of this paper as
    \begin{equation}
        \px^{\ast} = \underset{\px\geq 0}{\mathrm{argmax}}\,\underset{\nu, \, \lambda \geq 0}{\inf} \func{}\pasq{\px,\nu,\lambda}.\label{eq:px_max_inf_problem}
    \end{equation}
\end{definition}
%%%%%%%%%%%%%%%%%%%%%%%%% Results %%%%%%%%%%%%%%%%%%%%%%%%%
\section{results}
\subsection{Capacity-achieving input distribution}
Before constructing the solution of \eqref{eq:px_max_inf_problem}, we want to establish its uniqueness, concretely we will show in the following two lemmata that, for a given maximal cost $\cbar$ there is a unique tuple $\pa{\px^{\ast},\nu^{\ast},\lambda^{\ast}}$ extremizing $\func{}\pasq{\px,\nu,\lambda}$. We achieve this by demonstrating that $\func{}\pasq{\px,\nu,\lambda}$ is strictly concave in $\px$ and strictly convex in $\pa{\nu,\lambda}$.
\begin{lemma}[Strict concavity of mutual information in $\px$]
    \label{lem:convexity_px}
    Let $p_{X,\pa{\lambda,\nu}}^{\ast}\pa{x}$ be the solution to the problem $\underset{\px}{\max}\, \func{}\pasq{\px,\nu,\lambda}$ for a fixed $\pa{\nu,\lambda}$. Then
    \begin{enumerate}
        \item $\func{0}\pasq{\px,\nu}$ and $\func{}\pasq{\px,\nu,\lambda}$ are strictly concave in $\px$\label{concavity_MI},
        \item $p_{X,\pa{\lambda,\nu}}^{\ast}\pa{x} \neq p_{X,\pa{\lambda^{\prime},\nu^{\prime}}}^{\ast}\pa{x}$ if $\lambda\neq \lambda^{\prime}$\label{lbda_to_pX_injective} %the mapping from $\lambda^{\ast}$ to $\px^{\ast}$ is injective.  
    \end{enumerate}
\end{lemma}
\begin{IEEEproof}
    To prove \ref{concavity_MI}, we compute the Hessian of $\func{0}$ with respect to $\px$ and show that it is strictly negative definite.
    The proof relies on the boundedness of the support of $\px$, i.e., the PA constraint.
    The strict concavity of $\func{}$ follows from that of $\func{0}$ because they differ only by a term linear in $\px$.
    %; in the periodic case, only negative (not strict) definiteness is guaranteed.\\
    
    The property \ref{lbda_to_pX_injective} then follows from the fact that, as we will show in Appendix \ref{subsec:Detailed_proof_convexity_in_px}, in the functional for the constrained problem, \eqref{eq:px_max_inf_problem}, the extremizations over $\px$ and the Lagrange multipliers can be swapped and that $\lambda$ is coupled to $\px$ only by a term linear in $\px$. The details of the proof are presented in Appendix \ref{subsec:Detailed_proof_convexity_in_px}.
\end{IEEEproof}

\begin{lemma}[Uniqueness of $\lambda^{\ast}$]
\label{lem:chain_non_overlapping}
For fixed $\bar{c}\in\left(0,\bar{c}^{\ast}\right]$,
%let $\px^{\lambda}$ be the discrete probability distribution \eqref{eq:general_ansatz} with $x_j$ and $m_{j}\left(\lambda\right)$ as defined in (\ref{eq:def_pos_unconstr}) and (\ref{eq:def_masses_unconstr}).
there is a unique value of the Lagrange multiplier $\lambda=\lambda^{\ast}$ such that $\pa{\px^{\ast},\nu^{\ast}, \lambda^{\ast}}$ is a solution to the optimization problem \eqref{eq:px_max_inf_problem}.
\end{lemma}
\begin{IEEEproof}
    We again use that in \eqref{eq:px_max_inf_problem}, the extremizations of maximizing over $\px$ and building the infimum over the Lagrange multipliers can be swapped. Then, we show that the functional maximized over $\px$ is strictly convex in the Lagrange multipliers (cf. \cite[sec. 5.1.2.]{Boyd_convex_optimization_2004}). The uniqueness then follows from the fact that $\cbar$ is coupled to $\lambda$ only by a term linear in $\lambda$. We defer details to Appendix \ref{subsec:uniqueness_lambda_ast}.    
\end{IEEEproof}

In order to construct the solution to \eqref{eq:px_max_inf_problem}, we use the conditions that Smith has shown to be necessary and sufficient for $\px^{\ast}$ to be the capacity-achieving input distribution of a channel with additive noise and PA constraint \cite{smith_information_1971}. 
Even though he considers
Gaussian additive noise and a constraint on the second moment, i.e.
$c\left(x\right)=x^{2}$, his derivation of the following lemma holds
for arbitrary additive noise and arbitrary cost function.

\begin{figure}
\centering{\includegraphics[width=\figW]{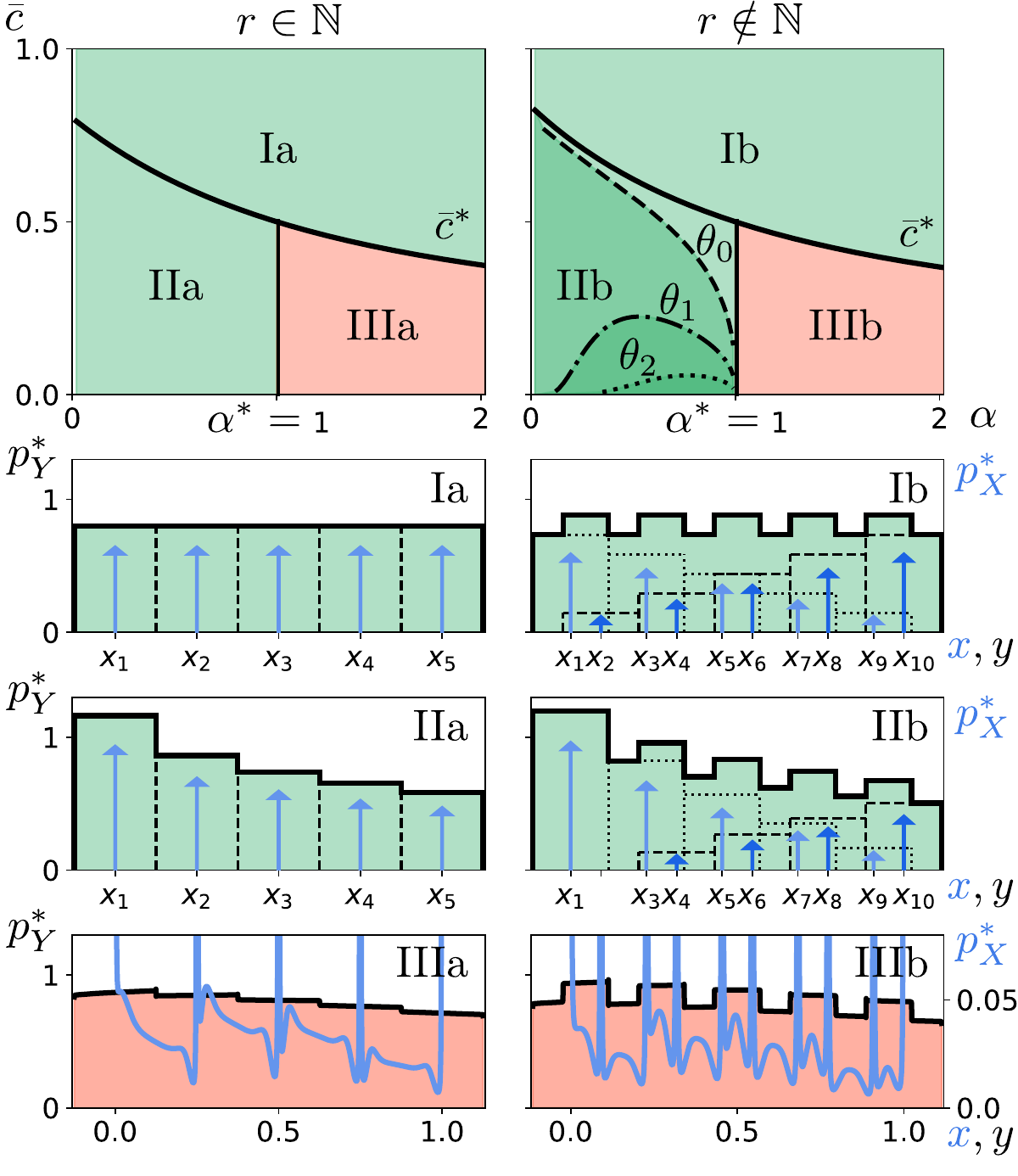}}
\caption{The different cases discussed in \thmref{main}.
In the left column $r\in\mathbb{N}$ ($r=4$) and in the right column $r\notin\mathbb{N}$ ($r=4.4$).
Top: Phase diagram in the $\alpha$-$\bar{c}$-plane.
Green and red background indicate $\px^{\ast}$ with discrete support and
support on the entire interval $[0,1]$, respectively.
Ia,b and IIa,b: discrete $\px^{\ast}$ with masses $m_j$ and positions $x_j$ indicated by the heights and the positions of the blue arrows (dark blue for j even). The corresponding $\pyx(y-x_j)$ is illustrated by dashed boxes in Ia/IIa and by dotted (j odd) and dashed (j even) boxes in Ib/IIb.
The black line is the resulting $\py^{\ast}$.
IIIa and IIIb: numerical result for $\px^{\ast}$ (blue) using the Blahut-Arimoto algorithm \cite{blahut_1972, arimoto_1972} and corresponding $\py^{\ast}$ in black.\vspace{-0.3cm}
\label{fig:sketch_integer_case}}
\end{figure}

%%%%%%%%%%%%%%%%%%%%% Smith's theorem %%%%%%%%%%%%%%%%%%%%%
\begin{theorem}[Optimality conditions; Smith, \cite{smith_information_1971}]
Let $C$ denote the channel capacity. Then, for an additive
channel with PA and a cost constraint of the form $\left\langle c\right\rangle \leq \bar{c}$,
the capacity-achieving input distribution $\px^{\ast}$ implicitly
defined by
\begin{equation}
C=\max_{\begin{array}{c}
\px \colon \\
\int_0^1 dx \, \px(x) = 1\\
\left\langle c\left(x\right)\right\rangle \leq\bar{c}
\end{array}}I\left(X;Y\right),\label{eq:channel_capacity}
\end{equation}
is unique and determined by the necessary and sufficient conditions
\begin{alignat}{2}
i\left(x;\px^{\ast}\right) & \leq I\left(\px^{\ast}\right)+\lambda\left(c\left(x\right)-\bar{c}\right) & \quad & \mathrm{for}\;\mathrm{all}\;x\in\left[0,1\right],\label{eq:ineq_constr}\\
i\left(x;\px^{\ast}\right) & =I\left(\px^{\ast}\right)+\lambda\left(c\left(x\right)-\bar{c}\right) & \quad & \mathrm{for}\;\mathrm{all}\;x\in S,\label{eq:eq_constr}\\
1&=\int_0^1 dx\,\px\pa{x}
\end{alignat}
where $S$ denotes the support of $\px^{\ast}$,
\begin{equation}
i\pa{x;\px} \coloneqq\int dy\,p_{\mathrm{N}}\pa{y-x}\log \left(\frac{p_{\mathrm{N}}\pa{y-x}}{\py\pa{y;\px}} \right)\label{eq:def_info_density}
\end{equation}
 is the marginal information density, and $I\left(\px\right)\coloneqq\int dx\,\px\left(x\right)\,i\left(x;\px\right)$
is the mutual information between $X$ and $Y$.
\end{theorem}
\begin{IEEEproof}
This follows from the condition that for $\px$ to maximize the mutual information in \eqref{eq:channel_capacity}, the functional derivative $\delta \func{}/\delta \px$ has to vanish inside the domain of allowed values for $\px\pa{x}$ - given that $\px$ is a probability, the allowed values are the $x$ values for which $\px\pa{x}>0$. For $\px\pa{x} = 0$, the functional derivative has to be negative. This yields the Karush-Kuhn-Tucker (KKT) equations \cite{Karush39, Kuhn51_481, Boyd_convex_optimization_2004}, which are equivalent to~\eqref{eq:ineq_constr} and \eqref{eq:eq_constr}. For details see \cite{smith_information_1971}, replacing the variance constraint $x^{2}$ by the general cost constraint $c\left(x\right)$.
\end{IEEEproof}
To state the main theorem concisely and for later reference, we first introduce the following definitions.
\begin{definition}
    We define $\px^{0}$ as the capacity-achieving input distribution in the absence of the cost constraint, or if the constraint is not active, i.e. $\lambda=0$.
    The corresponding output distribution we denote as $\py^{0}$.
    Furthermore, we define the critical expected cost $\bar{c}^\ast \coloneqq \Ex{c(x)}_{\px^{0\ast}}$ as the cost below which the cost constraint becomes active.
\end{definition}
\begin{notation}
In the parameter regimes in which the capacity-achieving input distribution is discrete, we parametrize it as
\begin{equation}
\px\pa{x}\coloneqq\sum_{j=1}^{N_{r}}m_{j}\,\delta\pa{x-x_{j}},\label{eq:general_ansatz}
\end{equation}
where the $N_r$ probability masses $m_{j}\geq0$ and their positions $x_{j}$ are the parameters that need to be determined.
\end{notation}
%\begin{remark}
%    The optimality conditions distinguish between points that are in the support, which have to meet the equality constraint \eqref{eq:eq_constr}, and those that are not, where only the inequality \eqref{eq:ineq_constr} has to hold.
%    This discontinuity leads to a different set of equations for an ansatz \eqref{eq:general_ansatz} with the assumption $m_{j^{\prime}}>0$ compared to an ansatz, where $m_{j^{\prime}}=0$ is assumed.
%    Hence, the approach to determine the parameters of the ansatz \eqref{eq:general_ansatz} involves two steps. Fist, we guess the number and the positions of the non-vanishing masses, i.e. the support of $\px$, and in a second step we determine the values of those masses.
%\end{remark}
%
\begin{definition}
If the inverse width of the noise is non-integer, i.e. $r\notin \mathbb{N}$, we will show that the noisy output $p_{\mathrm{N}}\left(y\mid x_{j}\right)$ and $p_{\mathrm{N}}\left(y\mid x_{j+1}\right)$ of two neighboring inputs $x_j$ and $x_{j+1}$ overlap.
Due to those overlaps, the output distribution is made of $2n+1$ consecutive constant segments, see \figref{sketch_integer_case} (Ib) and (IIb), whose heights are given by $\mhat{}$ (odd-indexed heights) and $\mbar{}$ (even-indexed heights) defined as
\begin{equation}
\begin{split}
\mhat{} & \coloneqq\pa{m_{1},\,m_{2}+m_{3},\,\ldots,\,m_{2n}+m_{2n+1},\,m_{2n+2}},\\
\mbar{} & \coloneqq\pa{m_{1}+m_{2},\,m_{3}+m_{4},\,\ldots,\,m_{2n+1}+m_{2n+2}},
\end{split}\label{eq:def_mhat_mbar}
\end{equation}
where
\begin{equation}
    n\coloneqq\floor{r} + 1
\end{equation}
is the integer part of the inverse width $r$ of the noise plus one.
The sum of the entries of $\mhat{}$ and $\mbar{}$ equal that of all masses, i.e.
\begin{equation}
\sum_{j=1}^{n+1}\mhat{j}=\sum_{j=1}^{n}\mbar{j}=\sum_{j=1}^{N_{r}}m_{j}=1.\label{eq:sum_constraint}
\end{equation}
We obtain the original probability masses via the back transform given by
\begin{equation}
m_{j}=\begin{cases}
\sum_{l=1}^{\left(j+1\right)/2}\mhat{l}-\sum_{l=1}^{\left(j-1\right)/2}\mbar{l} & \mathrm{if}\;j\;\mathrm{is}\;\mathrm{odd,}\\
\sum_{l=1}^{j/2}\left(\mbar{l}-\mhat{l}\right) & \mathrm{if}\;j\;\mathrm{is}\;\mathrm{even}.
\end{cases}\label{eq:def-masses-non-integer-case}
\end{equation}
\end{definition}
\begin{remark}
    Note that the back transform \eqref{eq:def-masses-non-integer-case} does not guarantee $m_j\geq 0$ even if all combined masses $\mhat{}$ and $\mbar{}$ are positive.
    Indeed, from a technical perspective, this is one of the major challenges.
\end{remark}
\begin{definition}
    To streamline index ranges, we define $\pasq{k:l}\coloneqq\pacu{i\in \mathbb{N}\mid k\leq i \leq l}$ and write $\pasq{l}$ if $k=1$.
\end{definition}
%
%\begin{definition}
%For convenience, we also introduce two types of subsets $\Mkl{k}$
%and $\Mkg{k}$ of the probability masses. We define
%\begin{align}
%\Mkl{k} & \coloneqq\begin{cases}
%\emptyset, & k=0\\
%\pacu{m_{2j-1}\mid j=1,\,\ldots,\,k}, & k>0
%\end{cases},\\
%\Mkg{k} & \coloneqq\pacu{m_{j}\mid j=2k+1,\,\ldots,\,2n+2},
%\end{align}
%for $k=0,\,\ldots,\,n$, so that $\Mkl{k}$ contains the odd labeled
%masses up to $m_{2k-1}$, and $\Mkg{k}$contains all masses starting
%from $m_{2k+1}$. In the special case $k=0$, all masses are in $\Mkg{0}$
%and $\Mkl{0}$ is empty. We denote the sums over the masses by
%\begin{align}
%    \mkl{k}&\coloneqq\sum_{m_j\in \Mkl{k}}m_j&\mathrm{and}& & \mkg{k}&\coloneqq\sum_{m_j\in \Mkg{k}}m_j.
%\end{align}
%The corresponding positions are contained
%in the subsets $\xkl{k}$ and $\xkg{k}$ defined as
%\begin{align}
%\xkl{k} & \coloneqq\begin{cases}
%\emptyset, & k=0\\
%\pacu{x_{2j-1}\mid j=1,\,\ldots,\,k}, & k>0
%\end{cases},\\
%\xkg{k} & \coloneqq\pacu{x_{j}\mid j=2k+1,\,\ldots,\,2n+2}.
%\end{align}
%\end{definition}

%%%%%%%%%%%%%%%%%%%%% main theorem %%%%%%%%%%%%%%%%%%%%%
\begin{theorem}[Main Theorem]
\label{thm:main}The capacity-achieving input distribution $\px^{*}$
of the additive uniform noise channel with peak amplitude and cost
constraint with cost function $c(x)$ as specified in \Cref{ass:cost_function}, has the following properties:
\begin{casenv}
% Case I
%%%%%%%%%
\item[I] (Oettli, \cite{oettli_capacity-achieving_1974}) \label{case:i}If the cost constraint is inactive (i.e. $\bar{c}\geq\bar{c}^{\ast}$),
then the capacity-achieving input distribution is given by \eqref{eq:general_ansatz} where 
\begin{align}
N_r & =\begin{cases}
n & \mathrm{if}\;r\in\mathbb{N}\\
2n& \mathrm{if}\;r\notin\mathbb{N}
\end{cases}\label{eq:npoints}
\end{align}
is the number of mass points with $n=\floor{r} + 1$. The mass locations $x_{j}$ and the masses $m_{j}$ are given
by
\begin{align}
x_{j} & =\begin{cases}
\frac{j-1}{n-1} & \mathrm{if}\;r\in\mathbb{N}\\
\frac{j-1}{2r} & \mathrm{if}\;r\notin\mathbb{N},\;j\;\mathrm{is}\;\mathrm{odd}\\
1-\frac{2n-j}{2r} & \mathrm{if}\;r\notin\mathbb{N},\;j\;\mathrm{is}\;\mathrm{even},
\end{cases}\label{eq:def_pos_unconstr}\\
m_{j} & =\begin{cases}
\frac{1}{n} & \mathrm{if}\;r\in\mathbb{N}\\
\frac{2n-\left(j-1\right)}{2n\pa{n+1}} & \mathrm{if}\;r\notin\mathbb{N},\;j\;\mathrm{is}\;\mathrm{odd}\\
\frac{j}{2n\pa{n+1}} & \mathrm{if}\;r\notin\mathbb{N},\;j\;\mathrm{is}\;\mathrm{even},
\end{cases}\label{eq:def_masses_unconstr}
\end{align}
where $j\in\rangeone{N_r}$. Thus, the support
of $\px^{\ast}$ is discrete and given by $S_{0}\coloneqq\left\{ x_{j}\mid j=1,\ldots,N_r\right\} $.
\figref{sketch_integer_case} (Ia and Ib) illustrate $\px^\ast$.
% Case new IIa
%%%%%%%%%%
\item[IIa] \label{case:iia}If the cost constraint is active ($\bar{c}<\bar{c}^{\ast}$), the
cost function $c(x)$ is concave ($\alpha\leq1$), and $r\in\mathbb{N}$, then the capacity-achieving input distribution is discrete with mass locations
as in (\ref{eq:def_pos_unconstr}) and masses given by
\begin{equation}
m_{j} =\frac{1}{z}e^{-\lambda^\ast c_{j}}, \quad z =\sum_{j=1}^{N_r}e^{-\lambda^\ast c_{j}},\label{eq:masses_integer_case}
\end{equation}
where $c_{j}\coloneqq c\pa{x_{j}}$, and there exists a $\lambda^\ast > 0$ such that the cost constraint is satisfied.
Thus, the support of $\px^{\ast}$ is given by $S_{0}$,
see \figref{sketch_integer_case} (IIa)% for an illustration of $\px^\ast$.
% Case new IIb
%%%%%%%%%
\item[IIb] \label{case:iib} If the cost constraint is active ($\bar{c}<\bar{c}^{\ast}$), the
cost function $c(x)$ is concave ($\alpha\leq1$), and $r\notin\mathbb{N}$, then
the capacity-achieving input distribution is discrete. Furthermore, there exist $n-1$ thresholds $0<\theta_{n-2}<\ldots<\theta_0<\bar{c}^*$ such that the support can be expressed as
\begin{align}
    S  = \begin{cases}
S_0 & \mathrm{if} ~ \bar{c} >\theta_0\\
S_k & \mathrm{if} ~ \bar{c}\in (\theta_k,\theta_{k-1}],~k\in\rangeone{n-2}\\
S_{n-1} & \mathrm{if} ~ \bar{c}\in (0,\theta_{n-2}]
\end{cases}.\label{eq:Sk_cases}
 \end{align}
%where $S_k = S_{k-1}\setminus \{x_{2k}\}$, $1\leq k\leq n$.
Here, $S_k=\xkl{k}\, \cup\, \xkg{k}$, where
\begin{equation}
    \xkl{k}\coloneqq \pacu{x_{2j-1}\mid j=1,\,\ldots,\,k},
\end{equation}
contains the odd labeled positions of \eqref{eq:def_pos_unconstr} up to $x_{2k-1}$, and
\begin{equation}
    \xkg{k} \coloneqq\pacu{x_{j}\mid j=2k+1,\,\ldots,\,2n},
\end{equation}
contains all positions of \eqref{eq:def_pos_unconstr} starting from $x_{2k+1}$.\\
For $m_j \in \Mkl{k}$, where $\Mkl{k}\coloneqq \pacu{m_{j}\mid x_j\in \xkl{k}}$, the masses are given by
\begin{equation}
    m_{j} =\frac{1}{z}e^{-\lambda^\ast c_{j}}, \quad z =\sum_{j=1}^{k}\frac{e^{-\lambda^\ast c_{2j-1}}}{\mkl{k}}.\label{eq:m_smaller}
\end{equation}
For $m_j \in \Mkg{k}$, where $\Mkg{k} \coloneqq\pacu{m_{j}\mid x_j \in \xkg{k}}$, the masses can be computed via \eqref{eq:def-masses-non-integer-case}, with the combined masses given by 
\begin{align}
\mhat{j} & =\frac{1}{\zhat}e^{-\lambda^\ast \dhat{j}}, & \zhat= & \sum_{j=k+1}^{n+1}\frac{e^{-\lambda^\ast \dhat{j}}}{\mkg{k}}\label{eq:m_hat_j_expl_general},\\
\mbar{\ell} & =\frac{1}{\zbar}e^{-\lambda^\ast \dbar{\ell}}, & \zbar= & \sum_{j=k+1}^{n}\frac{e^{-\lambda^\ast \dbar{j}}}{\mkg{k}},\label{eq:m_bar_j_expl_general}
\end{align}
where $j\in\range{k+1}{n+1}$ and $\ell\in\range{k+1}{n}$.
Moreover,
\begin{align}
    \mkl{k}&\coloneqq\sum_{m_j\in \Mkl{k}}m_j&\mathrm{and}& & \mkg{k}&\coloneqq\sum_{m_j\in \Mkg{k}}m_j,
\end{align}
with $\mkl{k}+\mkg{k}=1$, denote the sums of the masses of the different subsets.
The exponents $\dhat{j}$ and $\dbar{j}$, which are cumulative sums of the differences between the cost function evaluated at neighboring $x_j$, are defined in Appendix \ref{appendix_A_prerequisites}, eqs.~\eqref{eq:definition_dhat_general} and \eqref{eq:definition_dbar_general}.
In the special case $k=0$, $\xkl{0}$ contains all $x_j$ and $\xkg{0}=\emptyset$.\\
There exists a value $\lambda^\ast >0$ for the Lagrange multiplier such that the cost constraint is satisfied.
See \figref{sketch_integer_case} (IIb).
%\figref{sketch_integer_case} (IIb) illustrates $p_x^\ast$ for $k=1$, i.e. $\bar{c}\in(\theta_1,\theta_0]$. 
%\figref{sketch_labeling} illustrates $\px^{\ast}$ and the labeling of the masses and positions for $k=3$, i.e. $\bar{c}\in(\theta_3,\theta_2]$.
% CaseIII
%%%%%%%%%%%
\item[III] \label{case:iii}If the cost constraint is active ($\bar{c}<\bar{c}^{\ast}$) and
the cost function $c(x)$ is strictly convex on a finite interval $G\subseteq\pasq{0,1}$ ($\alpha > 1$), then $G$ is in the support of the capacity-achieving input distribution.
In particular, if $c\pa{x}$ is strictly convex on $\pasq{0,1}$, then $\px^{\ast}$ has support on the entire interval $[0,1]$,
see \figref{sketch_integer_case} (IIIa and IIIb).
\end{casenv}
\end{theorem}
The top row of \figref{sketch_integer_case} shows the positions of the different cases
of \thmref{main} in the phase diagram.\\
%%%%%%%%%%%%%%%%%% Proof Case I %%%%%%%%%%%%%%%%%%
% no cost constraint (Oettli)
\begin{IEEEproof}[Proof of Case I (Oettli)]
% In \cite{oettli} Oettli computed the capacity-achieving input distribution
% of additive channels with piecewise constant noise and with PA. In
% this regard, (\ref{eq:cahnnel_definition}) and (\ref{eq:channel_capacity})
% are the special case of completely constant noise, for which we denote
% the capacity-achieving input distribution as $p_{0,x}^{\ast}$. However, from
% the perspective of the cost function, our considerations
% are a generalization of Oettli's result because he does not assume
% any cost constraint or, equivalently, he chooses $\bar{c}>\bar{c}^{\ast}$,
% where $\bar{c}^{\ast}\left(\alpha,r\right)\coloneqq\int dx\,p_{0,x}^{\ast}\left(x\right)\,x^{\alpha}$
% is the cost of the unconstrained case.
The full proof is given in \cite{oettli_capacity-achieving_1974}; we will only give a few hints here: %, and since our results are based on it, we briefly review it here.
If $r\in\mathbb{N}$, the width of the blocks $\pyx\pa{y - x}$ is such that $r+1$ of these blocks can cover the interval $D_{Y}$ perfectly without overlaps or gaps.
In this configuration, the touching blocks form a uniform output distribution $\py^{0}\pa{y}=\frac{n-1}{n}\mathbf{1}_{-b<y<1+b}$, see \figref{sketch_integer_case} (Ia), which is known to maximize the output entropy $H\left(Y\right)\coloneqq-\int dy\,\int dx\,\pyx\pa{y - x}\log \py\pa{y}$ if $Y$ is restricted to an interval $D_{Y}$.
Since the conditional entropy $H\pa{Y\mid X}$ is solely determined by the distribution of the noise, maximizing $H\pa{Y}$ with respect to $\px$ is equivalent to maximizing $I\pa{X;Y}$.

If $r\notin \mathbb{N}$, the first step is to guess the correct positions of the discrete mass points.
In order to avoid diverging marginal information densities, it is necessary to put masses both on $x=0$ and $x=1$.
The next mass points are then placed on two interleaved grids each with a step size of $2b$, one of them containing $0$, the other $1$ (cf. \figref{sketch_integer_case} (Ib)).
Inserting this ansatz into the equality constraint \eqref{eq:eq_constr} yields equations for the values of the masses similar to that of the case $r\in \mathbb{N}$ but with the additional complication that each equation depends on multiple $m_j$.
They can nevertheless be explicitly resolved. 
Finally, one can show that the resulting
$\py^{0}\pa{y;\px^{0}}$ is $2b$-periodic within
the interval $D_{Y}\coloneqq\left[-b,1+b\right]$, which leads to a constant
$i(x;\px^\ast)=I(X;Y)$ and therefore, the inequality constraint \eqref{eq:ineq_constr} is also fulfilled (with equality for all $x\in \pasq{0,1}$).
\end{IEEEproof}
\begin{remark}
The uniform output distribution of the integer case $r\in \mathbb{N}$ can also be obtained in the limiting case $\rho:=r-\floor{r}\rightarrow0$ from above, where $\rho \in [0,1)$ measures the distance of $r$ to its last integer value.
Indeed in this limit, every pair of odd-indexed position and next even-indexed position converge to the same value, i.e. $\lim_{\rho\searrow 0}x_{j}=\lim_{\rho\searrow 0}x_{j+1}=\left(j-1\right)/\pa{2\pa{n-1}}$,
for $j=1,3,\ldots,N_{r}-1$ odd.
The corresponding masses $m_{j}$ and $m_{j+1}$ add up to $1/n$, we recover the integer case $r\in \mathbb{N}$.
\end{remark}

%%%%%%%%%%%%%%%%%% Figure: labeled masses %%%%%%%%%%%%%%%%%%
\begin{figure}
\centering{\includegraphics[width=0.95\figW]{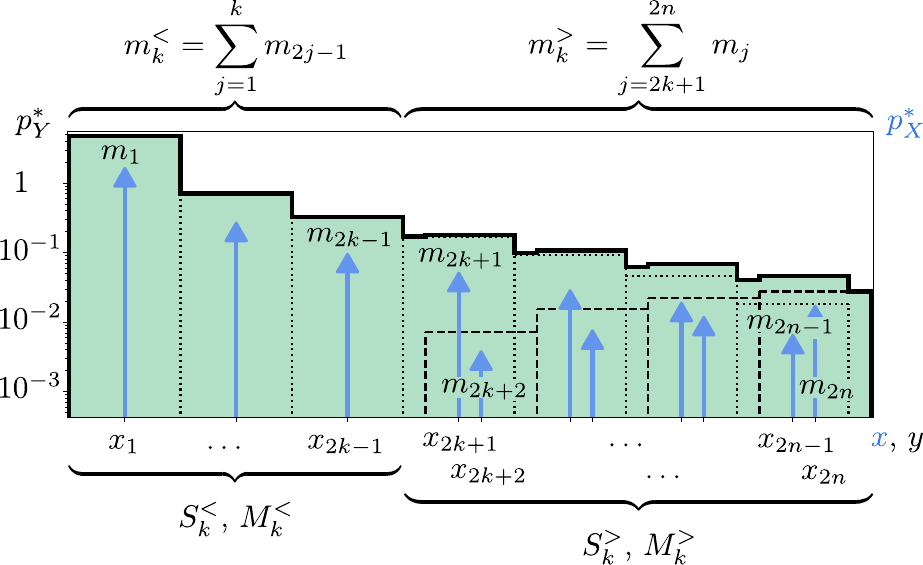}}
\caption{Illustration of the the positions and masses belonging to $\xkl{k}$, $\Mkl{k}$, $\xkg{k}$, and $\Mkg{k}$. Capacity-achieving input distribution $\px^{\ast}$ (blue arrows) and corresponding output distribution (black solid curve) for parameter values $r=6.2$, $\alpha=0.5$. The masses $m_2=m_4=m_6=0$ and $m_8>0$, so that the support is given by $S_k$ with $k=3$.\vspace{-0.3cm}
\label{fig:sketch_labeling}}
\end{figure}
\begin{figure}
\centering{\includegraphics[width=0.9\figW]{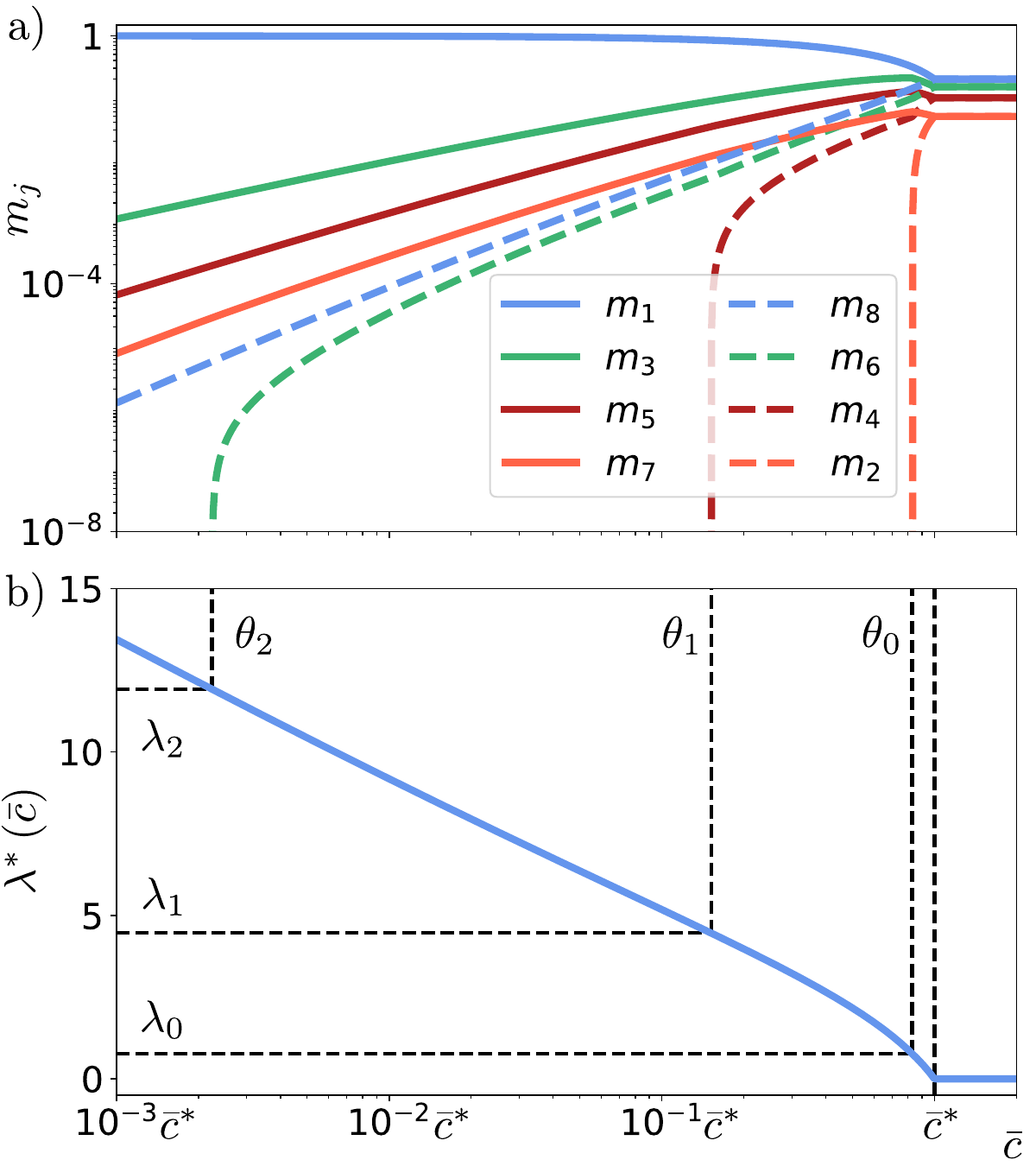}}
\caption{a) Masses $m_j$, $j=1,\,\dots,\,N_r$, as a function of $\cbar$. b) The Lagrange multiplier $\lambda^{\ast}$ as a function of $\cbar$. The values $\cbar=\theta_k$ (dashed vertical lines) denote the cost budget at which $m_{2k+2}$ vanishes and the support changes from $S_k$ to $S_{k+1}$ as the cost budget tightens. The corresponding values of the Lagrange multiplier are depicted by horizontal dashed lines. Here, $\alpha=0.5$ and $r=3.9$.\vspace{-0.3cm}
\label{fig:vanishing_masses}}
\end{figure}
%%%%%%%%%%%%%%%%% Proof Case IIa %%%%%%%%%%%%%%%%%
% alpha < 1, r in N
Our approach to prove Case IIa and IIb is constructive; for both cases we guess a solution and then prove that it fulfills the conditions \eqref{eq:eq_constr} and \eqref{eq:ineq_constr}. Concretely, we proceed along the following lines:
\begin{enumerate}
    \item Guess the support $S=\left\{x_{1},x_{2},\dots\right\}$ of $p_{X}$. 
    \item Insert this ansatz into the equality constraint \eqref{eq:eq_constr} and explicitly solve for $\pacu{m_{1},m_{2},\ldots}$. The resulting expression will depend on the Lagrange multiplier $\lambda$.\label{insertion_step}
    \item Check that the solution of \ref{insertion_step} also fulfills the inequality constraint \eqref{eq:ineq_constr}.
    \item Use convexity of the optimization problem to obtain that the Lagrange multiplier $\lambda^{\ast}$ is uniquely determined by $\cbar$.
\end{enumerate}
\begin{IEEEproof}[Proof of Case IIa]
For $\alpha\leq1$ and $r\in\mathbb{N}$, we will show that the input distribution \eqref{eq:general_ansatz} with $x_j$ and $m_j$ as defined in \eqref{eq:def_pos_unconstr} and \eqref{eq:def_masses_unconstr} meets the necessary and sufficient conditions \eqref{eq:ineq_constr} and \eqref{eq:eq_constr}.

The positions $x_j$ are such that the blocks $\pyx \pa{y - x_{j}}$ cover the domain of $Y$ without overlap or gaps within the interval $D_{Y}$. The
corresponding marginal information density \eqref{eq:def_info_density} is given by
$i\left(x_{j};\px^{\ast}\right)=-\log m_{j}$, so
that the equality constraint (\ref{eq:eq_constr}) evaluates to
\begin{equation}
-\log m_{j}=I+\lambda\left(c_{j}-\bar{c}\right),\;j=1,\ldots,N_r,\label{eq:eq_constr_integer_case}
\end{equation}
where $c_{j}\coloneqq c\left(x_{j}\right)$. The masses $m_{j}=m_{j}\left(\lambda\right)$,
and hence the corresponding probability distribution $\px^{\lambda}$,
depend on $\lambda$, but this dependence is omitted when clear from
context. Taking the difference between \eqref{eq:eq_constr_integer_case} evaluated at $j$ and $j+1$ yields
$n-1$ equations of the form
\begin{equation}
m_{j+1}=m_{j}e^{-\lambda\left(c_{j+1}-c_{j}\right)}\label{eq:diff_eqs_integer_case},
\end{equation}
$j\in\rangeone{n-1}$.
The $m_{j}$ are nonnegative and a decreasing series over $j$ because
$\lambda\geq0$ and $c_{j+1}-c_{j}>0$. Since $\sum_{j=1}^{N_r}m_{j}=1$,
the masses can be written in the form of (\ref{eq:masses_integer_case}).
%With the following lemma, we prove that a unique $\lambda^\ast$ satisfies the cost constraint.

To show that probability distributions with support $S_0$ satisfy the inequality constraint (\ref{eq:ineq_constr}), we use the following lemma.
\begin{lemma}[Piece-wise linearity of the marginal information density]
\label{lem:linear_i}
Let the positions $x_{j}$ be as defined in \eqref{eq:def_pos_unconstr}, and the corresponding masses $m_{j}\geq0$.
Then $i\pa{x;\px}$ is linear for $x\in\left[x_{j},x_{j+1}\right]$.
\end{lemma}
\begin{IEEEproof}
    The piecewise linearity follows from the fact that the marginal information density is given by the convolution of two piecewise constant functions, for illustration see \figref{linear-i}. For a full proof, featuring the values of the slopes, see Appendix \ref{subsec:proof_i_linear}.
\end{IEEEproof}
%%%%%%%%%%%%%%%%%%%%% Figure: piecewise lineas i %%%%%%%%%%%%%%%%%%%%%
\begin{figure}
\centering{\includegraphics[width=0.75\figW]{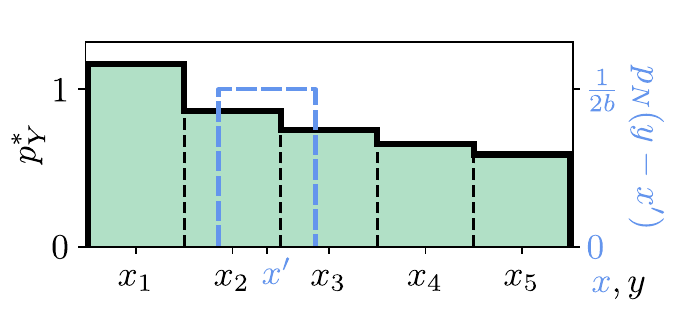}}
\caption{Illustration of the computation of the marginal information density.
In case of the uniform noise with density $\pyx\pa{y - x}$, the marginal information density evaluated at $x=x^{\prime}$ is given by the convolution of $\pyx\pa{y - x^{\prime}}$ (blue dashed box) with the output probability density $\py\pa{y}$ (black solid curve).}
\vspace{-0.3cm}
\label{fig:linear-i}
\end{figure}
\vspace{0.4\baselineskip}
%%%%%%%%%%%%%%%%%% Figure: lhs/rhs and p_X(x,alpha) %%%%%%%%%%%%%%%%%%
\begin{figure}
\centering{\includegraphics[width=1.\figW]{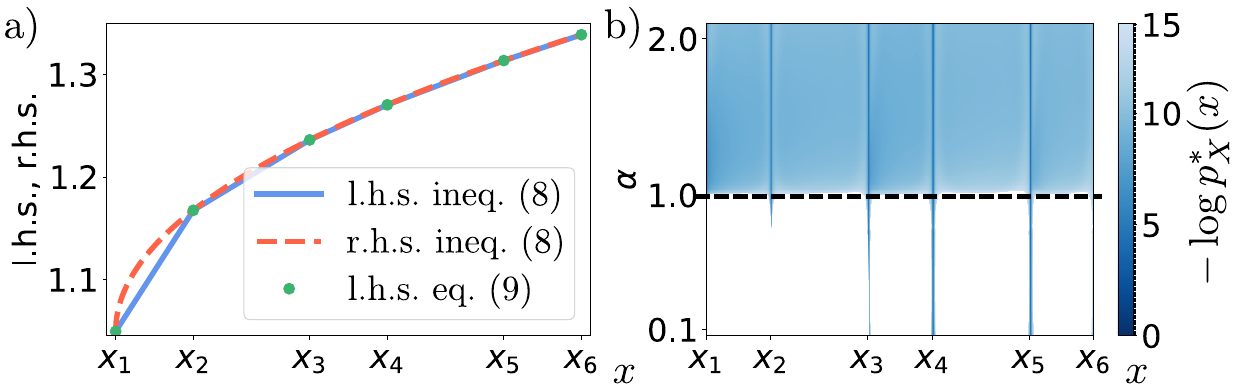}}
\caption{a) The r.h.s. and the l.h.s. of (\ref{eq:ineq_constr}), illustrating
the linear interpolation between the points of support, where (\ref{eq:eq_constr})
ensures equality.  Other parameters: $r=2.4$ and $\bar{c}=0.54<\bar{c}^\ast$. b) $\px^\ast\left(x\right)$ as a function of $\alpha$ obtained
numerically by means of the Blahut-Arimoto algorithm \cite{blahut_1972,arimoto_1972}, and ch.~7.2 of~\cite{dauwels_2006}.
For $\alpha\protect\leq1$, $\px$ is discrete and for $\alpha>1$, it has support on
the entire interval $[0,1]$. Other parameters: $r=2.4$ and $\bar{c}=0.35<\bar{c}^\ast$. \vspace{-0.3cm}
\label{fig:cost}}
\end{figure}
\vspace{0.4\baselineskip}
%&%&%&%&%&%&%
To conclude the proof of Case IIa, we note that, for every $\bar{c}\in\left[0,\bar{c}^{\ast}\right]$, 
\lemref{convexity_px} and \lemref{chain_non_overlapping} guarantee the
existence of a unique $\pa{\px^{\ast}, \nu^{\ast}, \lambda^{\ast}}$ that solves the equality constraint \eqref{eq:eq_constr}.
Therefore, the inequality constraint \eqref{eq:ineq_constr} is satisfied with equality at $x_{j}$,
$j\in\rangeone{N_r}$. Moreover, by \lemref{linear_i}, on the l.h.s. of (\ref{eq:ineq_constr}), $i\left(x;\px^{\ast}\right)$ increases linearly between $x_{j}$
and $x_{j+1}$. Due to the concavity of the cost function, the r.h.s of \eqref{eq:ineq_constr} is concave. Thus, the inequality constraint is also satisfied
for all the intermediate points $x\in\left(x_{j},x_{j+1}\right)$, see \figref{cost} a).
Hence, $\px^{\lambda^{\ast}}$ is the capacity-achieving input distribution $\px^{\ast}$ and its support is $S_0$, i.e. that of the unconstrained case.
This proves Case IIa. %\caseref{iii}.
\end{IEEEproof}
%%%%%%%%%%%%%%%%%% Proof Case IIb %%%%%%%%%%%%%%%%%
\begin{IEEEproof}[Proof of Case IIb]
We outline the proof here, while Appendix \ref{sec:proof_case_IIb} provides a detailed version.

In the non-integer case, i.e. $r\notin\mathbb{N}$, we proceed in three steps corresponding to the three cases in (\ref{eq:Sk_cases}).
First, in \emph{Step A}, we focus on $\bar{c} > \theta_0$, where the support is given by $S_0$.
Similarly to our proof of Case IIa, we derive the form of the capacity-achieving distribution and show that it
satisfies the necessary and sufficient conditions (\ref{eq:ineq_constr}) and (\ref{eq:eq_constr}).

%For $\cbar \leq \theta_0$ we will show that mass points will progressively disappear as the cost budget $\cbar$ decreases.
%First, when $\cbar$ reaches the threshold $\theta_0$, the mass $m_2$ equals zero.
%Indeed, for $\cbar<\theta_0$, the ansatz \eqref{eq:general_ansatz} with support $S_0$ results in negative masses when using the back transform \eqref{eq:def-masses-non-integer-case}.
%Dropping $x_2$ from the support will lead to a valid solution for $\px^{\ast}$, until $m_4=0$ when $\cbar=\theta_{1}<\theta_{0}$.
%Consecutively, all masses with even labels vanish until $m_{2n+2}$ is the only even labeled mass that continues to exist. This motivates the definitions of the supports $S_k=S_{k-1}/x_{2k+2}$, $k>0$.

As the cost budget $\cbar$ decreases below $\theta_0$, we will show that mass points vanish successively.
At $\cbar=\theta_0$, the mass $m_2$ is zero.
For $\cbar<\theta_0$, the ansatz \eqref{eq:general_ansatz} with support $S_0$ becomes infeasible because the back transform \eqref{eq:def-masses-non-integer-case} yields a negative value for $m_2$.
Removing $x_2$ from the support splits the set of all mass points into the two subsets $\Mkl{1}$ (to the left of $x_2$) and $\Mkg{1}$ (to the right of $x_2$), cf. \figref{sketch_labeling}.
In the right set, the equations defining the values of the mass points are as in Step A, only that the normalization is reduced by the weight of the leftmost mass point.
This approach yields a feasible solution $\px^\ast$ for $\cbar\in(\theta_1,\theta_0]$.
%The masses $m_{j>2}$
Upon further reducing $\cbar$, the same phenomenon reproduces so that 
at $\cbar=\theta_1<\theta_0$, the next even-indexed mass vanishes, i.e. $m_4=0$.
Proceeding in this way, the even-indexed masses disappear in order, see \figref{vanishing_masses} a).
The equations for the remaining odd-indexed masses $m_j\in \Mkl{k}$ are equivalent to those of \textit{Step A}, where $m_{j}$ depends only on $x_{j}$.
Finally, $m_{2n}$ is the only remaining even-indexed mass.
This motivates the recursive supports $S_k=S_{k-1} \backslash\{x_{2k+2}\}$, $k>0$.

Therefore, in \emph{Step B}, we show iteratively that these choices of the support yield valid solutions for the capacity-achieving input distribution on the intervals $\cbar \in (\theta_k, \theta_{k-1}]$ by demonstrating that the solutions we have guessed fulfill also the inequality constraint \eqref{eq:ineq_constr}.
Finally, in \emph{Step C}, we show that on the remaining interval $\cbar\in(0, \theta_{n-2}]$, the support is indeed given by $S_{n-1}$.
\end{IEEEproof}
%%%%%%%%%%%%%%%%%%%%%%%%%%%% Proof Case III %%%%%%%%%%%%%%%%%%%%%%%%%%%%%
%%%%%%%%%%%%%%%%%%%%%%%%%%%%%%%%%%%%%%%%%%%%%%%%%%%%%%%%%%%%%%%%%%%%%%%%%
\begin{figure}
\centerline{\includegraphics[width=0.75\figW]{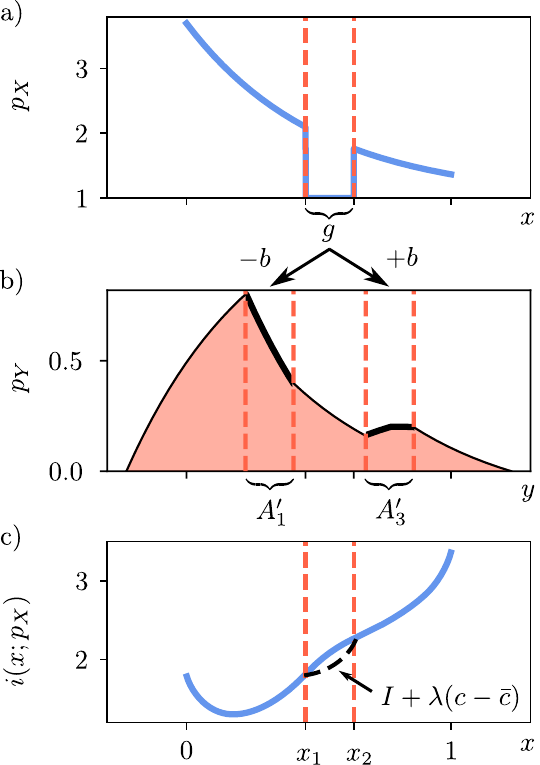}}
\caption{Illustration of the proof of Case III (convex cost function). a) Probability distribution $\px$ with a gap $g=(x_1,x_2)\notin S$. b) Corresponding output distribution $\py$ which is monotonically decreasing on $A^{\prime}_{1}$ and monotonically increasing on $A^{\prime}_{3}$. c) The resulting marginal information density is concave on the interval $g$. This violates the the inequality constraint \eqref{eq:ineq_constr} whose r.h.s. is convex (black dashed curve) due to the convex cost function $c\pa{x}$.
\vspace{-0.3cm}
\label{fig:proof_case_III}}
\end{figure}
%%%%%%%%%%%%%%%%%%%%%%%%%%%%%%%%%%%%%%%%%%%%%%%%%%%%%%%%%%%%%%%%%%%%%%%%%
% alpha > 1
\begin{IEEEproof}[Proof of Case III]
First, we note that it is impossible that there exists $\epsilon>0$ such that $\px^{\ast}\left(x\right)=0$ for all $x\in\left[0,\epsilon\right]$ because otherwise $\py^{\ast}\pa{y}=0$ for $y\in\left[-b,-b+\epsilon\right]$. 
With $i\left(x;\px^{\ast}\right)$ in \eqref{eq:def_info_density}, it would follow that $i\left(x,\px^{\ast}\right)\rightarrow\infty$, which would violate the inequality constraint \eqref{eq:ineq_constr}.
For the same reason, $\px^{\ast}=0$ on the interval $\left[1-\epsilon,1\right]$ and gaps of width $d\geq2b$ in $S$ are incompatible with (\ref{eq:ineq_constr}). 

Now, we will prove by contradiction that on an interval $G\subseteq\pasq{0,1}$, where the cost function is strictly convex, the support $S$ also cannot have finite gaps of width smaller than $2b$.
Define the interval $g\coloneqq\left(x_{1},x_{2}\right)$ of finite measure, where $0<x_{1}<x_{2}<1$.
Assume that $x_{1},x_{2}\in S$ and $g\not \subseteq S$, see \figref{proof_case_III} a).
Then, the equality constraint \eqref{eq:eq_constr} has to be satisfied at $x_{1}$ and $x_{2}$, and the inequality constraint \eqref{eq:ineq_constr} between these two points.
If $\alpha>1$, the r.h.s. of \eqref{eq:ineq_constr} has a strictly convex shape, which, as we will show, cannot be matched by the l.h.s. of \eqref{eq:ineq_constr}. To see this, let $x_\beta \coloneqq\left(1-\beta\right)x_{1}+\beta\, x_{2}$,
$\beta\in\left[0,1\right]$, to interpolate between $x_{1}$ and $x_{2}$. Now, $i\left(x_\beta;\px^{\ast}\right)$ is defined as the integral of $f(y) \coloneqq-\log[2b\,\py^*(y)] / (2b)$ over $[x_\beta-b,x_\beta+b]$. We split this set into three subsets
$A_1 \coloneqq [x_\beta-b,x_2-b]$, 
$A_2 \coloneqq [x_2-b,x_1+b]$, and 
$A_3 \coloneqq [x_1+b, x_\beta+b]$. Note that $|A_1|=(1-\beta)(x_2-x_1)$ and $|A_3|=\beta(x_2-x_1)$. In addition, we define the left enlargement of $A_1$ as $A_1' = [x_1-b,x_2-b]$, with $|A_1'|=x_2-x_1$, see \figref{proof_case_III} b). Due to the gap, $\py^{\ast}\left(y\right)=\frac{1}{2b}\int_{y-b}^{y+b}dx\,\px^{\ast}\left(x\right)$
is a decreasing function of $y$ on the set $A_1'$, which implies that $f(y)$ is increasing and, due to the left enlargement of $A_1$, we have
\begin{equation}
    \frac{1}{|A_1'|} \int_{A_1'} dy \, f(y) \leq \frac{1}{|A_1|} \int_{A_1} dy \, f(y).
\end{equation}
Similarly, we can define the right enlargement $A_3' = [x_1+b,x_2+b]$ of $A_3$ with $|A_3'|=x_2-x_1$. Since $f(y)$ is a decreasing function on $A_3'$ due to the gap, we obtain as before
\begin{equation}
    \frac{1}{|A_3'|} \int_{A_3'} dy\,f(y) \leq  \frac{1}{|A_3|} \int_{A_3} dy\,f(y) .
\end{equation}
Using the two inequalities above, we obtain
\begin{align}
    &i((1-\beta)\,x_1 + \beta\,x_2;\px^*) = \int_{A_1 \cup A_2 \cup A_3}dy\,f(y) \\
    \geq &(1-\beta) \int_{A_1'} dy\,f(y) + \int_{A_2} dy\,f(y) + \beta \int_{A_3'} dy\,f(y) \\
    = &(1-\beta)\, i(x_1;\px^*) + \beta \, i(x_2;\px^*),
\end{align}
using that $i(x_1;\px^*)$ and $i(x_2;\px^*)$ are the integrals of $f(y)$ over $A_1' \cup A_2$ and $A_2 \cup A_3'$, respectively.
This shows that $i\left(x,\px^{\ast}\right)$ is 
of concave shape.
Since we have equality at $x_1$ and $x_2$, this shows that the inequality constraint \eqref{eq:ineq_constr} is violated for $x\in g$, see \figref{proof_case_III} c).
Therefore, the only way to satisfy the inequality constraint on $G$ is to have $G\subseteq S$.
\end{IEEEproof}
\vspace{0.4\baselineskip}
\figref{cost} b) shows the transition from discrete to full support of
$\px^\ast$ for the example $c\pa{x}=x^{\alpha}$ when $\alpha$ crosses 1.
%%%%%%%%%%%%%%%%%%%%%%%%%%%%%%%%%%%%%%%%%%%%%%%%%%%%%%%%%%%%%%%%%%%%%%%%%
\subsection{Capacity}
\begin{definition}
    We denote the vector containing all masses of the discrete probability distribution as defined in \eqref{eq:general_ansatz} as
    \begin{equation}
        m\coloneqq \pa{m_1,\,\ldots,\,m_{N_r}}.
    \end{equation}
\end{definition}
\begin{definition}
The entropy of a discrete probability distribution with a vector of probability masses $m$ as defined above, we denote as
\begin{equation}
    H\pa{m}\coloneqq-\sum_{j=1}^{N_r} m_j \log m_j.
\end{equation}  
\end{definition}
%%%%%%%%%%%%%%%%%%%%% Figure: capacity %%%%%%%%%%%%%%%%%%%%%
\begin{figure}
\centering{\includegraphics[width=0.8\figW]{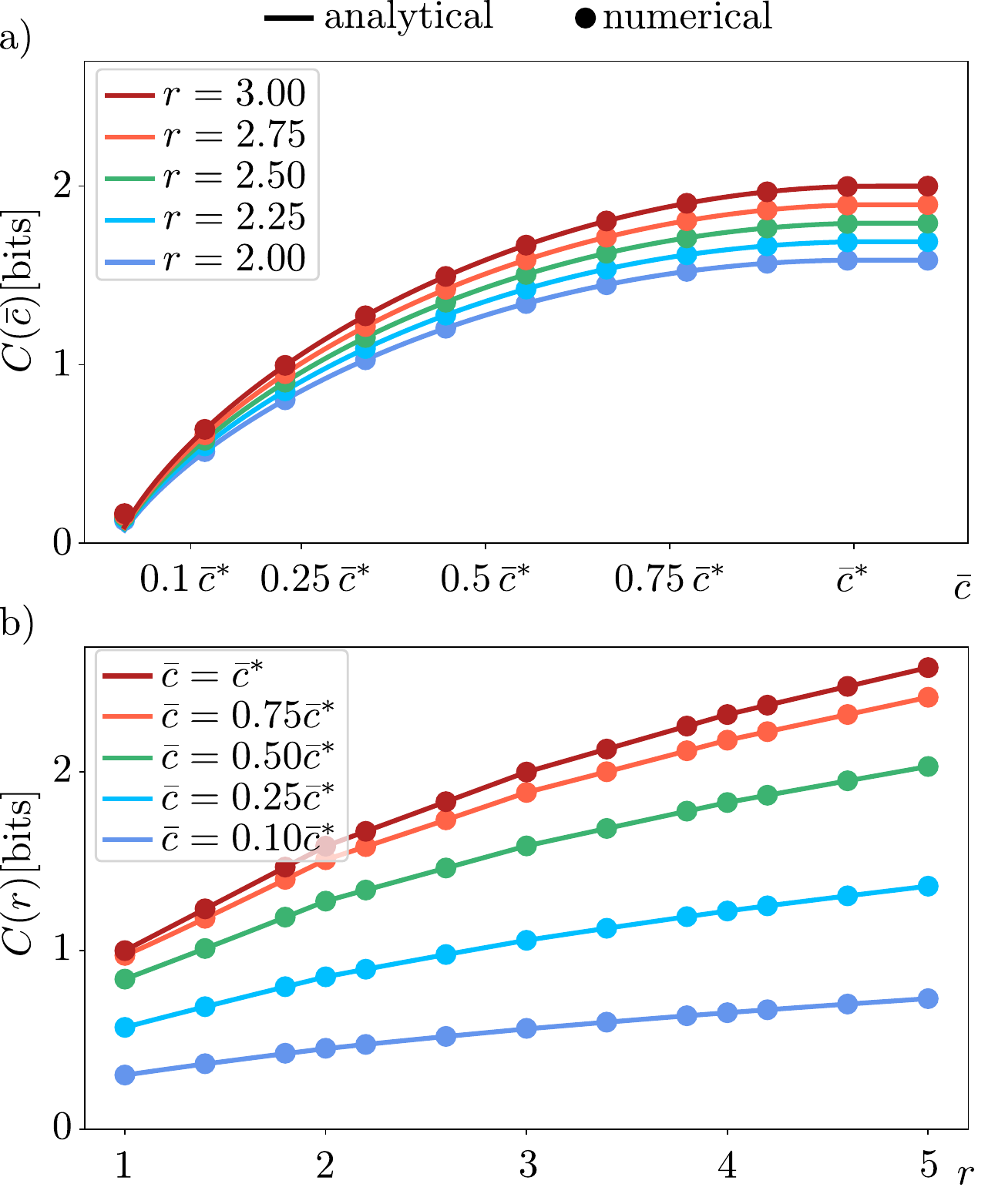}}
\caption{a) The capacity measured in bits as a function of the cost budget $\cbar$ for different values of the inverse noise width $r$. b) The capacity as a function of $r$ for different values of $\cbar$. In both panels, solid lines denote the analytical result presented in \thmref{capacity} and the dots indicate numerical results using the Blahut-Arimoto algorithm \cite{blahut_1972,arimoto_1972}, and ch.~7.2 of~\cite{dauwels_2006}.
All curves use the cost function $c\pa{x}=x^{\alpha}$ with $\alpha=0.7$.}
\vspace{-0.3cm}
\label{fig:capacity}
\end{figure}
\vspace{0.4\baselineskip}
%%%%%%%%%%%%%%%%%%%%% Theorem: capacity %%%%%%%%%%%%%%%%%%%%
\begin{theorem}
    The capacity of the additive uniform noise channel with peak amplitude constraint and a cost constraint with a concave cost function as specified in \Cref{ass:cost_function}, has the following properties:
    \begin{casenv}
    \item[I] If the cost constraint is inactive (i.e. $\bar{c}\geq\bar{c}^{\ast}$) and $r\in \mathbb{N}$, then the capacity is given by
    \begin{align}
    C &= 
    \begin{cases}
        \log\pa{n} & \mathrm{if}\;r\in\mathbb{N},\\
        \rho\log\pa{n+1} + \pa{1-\rho}\log\pa{n} & \mathrm{if}\;r\notin\mathbb{N},\label{eq:capacity_oettli}
    \end{cases}
    \end{align}
    where $\rho = r-\floor{r}$ and $n=\floor{r}+1$.
    \item[IIa] If the cost constraint is active ($\cbar<\cbar^{\ast}$), the cost function is concave ($\alpha\leq1$), and $r\in\mathbb{N}$, then the capacity is given by
    \begin{equation}
        C=H\pa{m},
    \end{equation}
    where the entries $m_j$ of $m$ are given by \eqref{eq:masses_integer_case}.
    \item[IIb] If the const constraint is active ($\cbar<\cbar{\ast}$), the cost function is concave ($\alpha\leq1$), and $r\notin\mathbb{N}$, then the capacity is given by
    \begin{equation}
        C=\rho H\pa{\mhat{}} + \pa{1-\rho} H\pa{\mbar{}},\label{eq:capacity_IIb}
    \end{equation}
    where the combined masses $\mhat{}$ and $\mbar{}$ are given by \eqref{eq:m_smaller}-\eqref{eq:m_bar_j_expl_general}.
    \end{casenv}\label{thm:capacity}
\end{theorem}
\begin{IEEEproof}[Proof of Case I]
    If $r\in\mathbb{N}$, there is no overlap between the noisy outputs $\pyx\pa{y - x_{j}}$ and $\pyx\pa{y - x_{j+1}}$ of two neighboring inputs $x_j$ and $x_{j+1}$, no confusion about the input symbol can arise (i.e. the conditional entropy is zero), so that the channel with uniform noise and PA constraint is equivalent to a noiseless channel with discrete input symbols $x_j$.
    Since all $n$ probability masses are equal and given by $1/n$, we obtain
    \begin{equation}
        C=-\sum_{j=1}^{n}\frac{1}{n}\log\pa{\frac{1}{n}}=\log\pa{n}.
    \end{equation}
    
    If $r\notin \mathbb{N}$, we use
    \begin{equation}
        I\pa{Y;X}=h\pa{Y}-h\pa{Y\mid X}, \label{eq:mutual_info_as_entropy}
    \end{equation}
    where $h\pa{Y}$ is the differential entropy of the continuous random variable $Y$, defined as
    \begin{equation}
        h\pa{Y}\coloneqq -\int dy\,\py\pa{y}\log \py\pa{y},
    \end{equation}
    and $h\pa{y\mid x}$ is the differential conditional entropy, defined as
    \begin{equation}
        h\pa{Y\mid X}\coloneqq -\int\int dx\,dy\,\pyx\pa{y - x}\,\px\pa{x}\log \pyx\pa{y - x}.
    \end{equation}
    Due to the additive noise, the conditional entropy is independent of $x$ and given by
    \begin{equation}
        h\pa{Y\mid X}=\log\pa{2b}.\label{eq:noise_etropy}
    \end{equation}
    The output distribution corresponding to the capacity-achieving input distribution
    $\px^{0}$ is given by the piece-wise constant function
    \begin{align}
        \py^{0}\pa{y}&=
        \begin{cases}
            \frac{r}{n+1}, &y\in\bigcup_{j=1}^{n+1} [\frac{2j-3}{2r},\,\frac{2\pa{j+\rho}-3}{2r}),\\
            \frac{r}{n}, &y\in\bigcup_{j=1}^{n} [\frac{2\pa{j+\rho}-3}{2r},\, \frac{2j-1}{2r}),
        \end{cases}
    \end{align}
    cf. \figref{sketch_integer_case} (Ib).
    The total length of the intervals, where $\py\pa{y}=r/\pa{n+1}$, is $\pa{n+1}\rho/r$ and the total length of the intervals, where $\py\pa{y}=r/n$, is $n\pa{1-\rho}/r$.
    Therefore,
    \begin{align}
        h\pa{Y}=&-\frac{\pa{n+1}\rho}{r}\,\frac{r}{n+1}\log\frac{r}{n+1}\nonumber\\
        &\;-\frac{n\pa{1-\rho}}{r}\,\frac{r}{n}\log\frac{r}{n}\\
        =& \rho \log \pa{n+1} + \pa{1-\rho}\log \pa{n} + \log 2b,
    \end{align}
    from which we obtain the second line of \eqref{eq:capacity_oettli} with the help of \eqref{eq:mutual_info_as_entropy} and \eqref{eq:noise_etropy}.
\end{IEEEproof}
\begin{IEEEproof}[Proof of Case IIa]
    Again, since $r\in\mathbb{N}$, there are no overlaps between the noisy outputs $\pyx\pa{y - x_{j}}$ and $\pyx \pa{y - x_{j+1}}$ of two neighboring inputs $x_j$ and $x_{j+1}$, so that the channel capacity is that of a discrete channel without noise and probability masses $m_j$ as defined in \eqref{eq:masses_integer_case}.
\end{IEEEproof}
\begin{IEEEproof}[Proof of Case IIb]
    In this case, the output distribution is given by
    \begin{align}
    \py^{\ast}\pa{y;\px^{\ast}}&=
        \begin{cases}
            r\,\mhat{j}, &y\in [\frac{2j-3}{2r},\,\frac{2\pa{j+\rho}-3}{2r}),\\
            r\,\mbar{j}, &y\in [\frac{2\pa{j+\rho}-3}{2r},\, \frac{2j-1}{2r}),
        \end{cases}
    \end{align}
    cf. \figref{sketch_integer_case} (IIb), where $j\in\rangeone{n+1}$ in the first line and $j\in\rangeone{n}$ in the second.
    In this expression we used that if $\cbar\in(\theta_{k},\theta_{k-1}]$, the even-indexed masses $m_{2},\,\ldots,\,m_{2k}$ vanish so that $\mhat{j}=\mbar{j}=m_{2j-1}$ for $j\in \rangeone{k}$, where $m_{2j-1}$ is given by \eqref{eq:m_smaller}.
    For $j\in\range{k+1}{N_r}$, the combined masses $\mhat{}$ and $\mbar{}$ are given by \eqref{eq:m_hat_j_expl_general} and \eqref{eq:m_bar_j_expl_general}.
    The differential entropy of the channel outpu can be written as
    \begin{align}
        h\pa{Y}=&-\sum_{j=1}^{n+1} r\,\mhat{j} \frac{\rho}{r}\log \pa{r\,\mhat{j}}\nonumber\\
        &\;-{}\sum_{j=1}^{n} r\,\mbar{j} \frac{1-\rho}{r}\log \pa{r\,\mbar{j}}\\
        =&-\rho \sum_{j=1}^{n+1} \mhat{j}\log \mhat{j}\nonumber\\
        &\;{}- \pa{1-\rho}\sum_{j=1}^{n} \mbar{j}\log \mbar{j} + \log 2b.
    \end{align}
    Using \eqref{eq:mutual_info_as_entropy} and \eqref{eq:noise_etropy}, yields \eqref{eq:capacity_IIb}.
\end{IEEEproof}
Panel a) of \figref{capacity} shows the capacity as a function of $\cbar$ and panel b) shows it as a function of $r$.
\begin{remark}
    For $r\notin \mathbb{N}$, the capacity is a convex combination of two capacities $H\pa{\mhat{}}$ and $H\pa{\mbar{}}$.
    Thus, the capacity is equivalent to that of a noiseless multiple-access channel with time sharing, where for a fraction $\rho$ of the time, the first code book with the $n+1$ probability masses $\mhat{}$ is used and for the rest of the time the code book with the $n$ probability masses $\mbar{}$ is used.
    Note that for $\cbar<\cbar^{\ast}$, the masses $\mhat{}$ and $\mbar{}$ are also functions of the fraction $\rho$. 
\end{remark}

%%%%%%%%%%%%%%%%%%%%%%%%%%%%%% Discussion %%%%%%%%%%%%%%%%%%%%%%%%%%%%%%%
\subsection{Discussion}
In this article, we have computed the capacity-achieving input distribution and the capacity for the uniform noise channel with peak amplitude (PA) constraint and average cost constraint.

% transition from discrete to continuous support
We have identified two mechanisms by which the capacity-achieving input distribution transitions from a purely discrete support to a (partly) continuous support.
(i) If the cost constraint is active, varying the cost function from a strictly concave shape (e.g., $c(x)=x^{\alpha}$ with $\alpha<1$) to a (partly) convex shape (e.g., $\alpha>1$) causes the support to contain a continuous component.
(ii) For a strictly convex cost function, decreasing the budget $\cbar$ past the critical level $\cbar{^\ast}$ likewise yields a continuous component.

% simpler proof?
Remarkably, when the cost function is strictly concave and hence the capacity-achieving input distribution is discrete, the possible positions of the mass points cannot be at other locations than the ones given by the support $S_0$ of the problem without cost constraint, i.e. $S\subseteq S_0$.
This holds true independently of the details of the shape of the cost function and independently of the maximal cost $\bar{c}$ (even though some mass points can vanish depending on $\cbar$ and the shape of the cost function).
This observation might hint towards a potentially simpler proof of the main theorem by using a generalization of the implicit function theorem. %Loosely speaking, by showing that $dx_j/d\bar{c} = 0$ for $\bar{c}\in (\theta_j,\theta_{k-1}]$.

% summary of the capacity
If the capacity achieving distribution is discrete, then the capacity of the channel is equivalent to that of a noiseless channel. If the width of the uniform noise is adjusted so that the noisy outputs of the discrete codewords cover the domain of the channel output without any overlaps, then the capacity is given by the entropy of the probabilities of the code words. In the other cases, where overlaps occur, the capacity is a convex combination of two entropies that interpolates between the previous and the next integer case.

% comparison with other work
This study can be seen as an extension of the work of Oettli \cite{oettli_capacity-achieving_1974}, since we consider an additional (tunable) cost constraint, which is the key ingredient that enables the phase transition between continuous and discrete capacity-achieving input distribution. This study also differs in two ways from the work of Tchamkerten \cite{tchamkerten_discreteness_2004}. First, we derive necessary and sufficient conditions (and not only sufficient conditions) for the emergence of discreteness for the capacity-achieving input distribution. Secondly, we consider an additive channel with bounded noise instead of unbounded noise.

There are a host of applications of methods to find capacity-achieving input distributions, many of them in biology: one can, e.g., think about the optimal shape of the input to a neuron or, as Witteveen et al. have done in a recent study, consider the optimal signal duration in a signaling pathway impinging on the gene expression of a downstream cell \cite{Witteveen25_arxiv}. It was found there that discrete input distributions maximize the shared information between input and output. Similarly, in biological neural networks it is found on a phenomenological level that signals are normally discrete - in fact, most neurons communicate by quasi-discrete events in time, known as action potentials or spikes \cite{Rieke97}. 
On a more theoretical level, also identifying the optimal prior for a statistical model is technically equivalent to finding a capacity-achieving input distribution and here as well, the input distribution is often found to be discrete \cite{Mattingly18_1760}.

More generally, we add to a series of works in which it was observed that discrete input distributions emerge in many cases under the objective to maximize the mutual information between input and output \cite{Fix78_704, Rose94_1939}. %TK: One of the oldest references for this seems to be Färber 1967, however, I could not find it online
The optimizing probability distributions in these cases are often found numerically, making it difficult to nail down the origin of their discreteness. Potentially, our fully analytical inroad could therefore yield deeper insights also in these cases.

% extensions
The present work could be extended in several directions. First, the results might extend to more general cost functions. In numerical simulations, we find that the capacity-achieving input distribution is discrete whenever a finite partition of the PA interval $[0,1]$ exists such that the cost function is concave on all subintervals. For any other PA interval, the problem can be reduced to the previous case by reparametrizing the cost function.

The present work could be extended in several directions. First, the current noise assumption (i.e. uniform noise) could be relaxed to a wider class of additive noises.
For example, if we would assume the noise density to be given by $p_N(N) \propto \exp(-|N/N_0|^\gamma)$, we would recover the uniform noise ($\gamma\rightarrow\infty$), the Gaussian noise ($\gamma = 2$) or the Laplace noise ($\gamma = 1$).
It remains to be investigated whether similar phase transitions occur when $\gamma < \infty$.
A second extension could be to replace the $x=1$ peak amplitude constraint with a softer constraint, e.g. $c(x)= x^\alpha + x^\beta, \forall x\leq 0$ and $\beta\geq 0$. The present PA corresponds to $\beta\rightarrow\infty$, whereas its absence would correspond to $\beta = 0$. This absence of PA could also be approached within the present framework in the limit of $\bar{c}\rightarrow 0$ and $r\rightarrow \infty$. This extension could help us determine to what extent the hardness of the constraint leads to the discrete support of the capacity-achieving input distribution. %
Another extension could be to consider a generalization of the capacity problem in higher dimensions where the input is restricted to a $L_1$ ball, analogously to the $L_2$ ball constraint for the additive vector Gaussian channel \cite{shamai_capacity_1995,dytso_capacity_2019,eisen_capacity-achieving_2023}.

\appendices
%%%%%%%%%%%%%%%%%%%%%%%%%%%%%% Appendix A %%%%%%%%%%%%%%%%%%%%%%%%%%%%%%%
\section{Proof of the uniqueness of the solution to the optimization problem \eqref{eq:px_max_inf_problem}}
\label{sec:appendixA}

%\subsection{Uniqueness of the extremizing $\px$ and $\lambda$}

\subsection{Proof of \Cref{{lem:convexity_px}} (Strict concavity of mutual information in $\px$)}
\label{subsec:Detailed_proof_convexity_in_px}
    To prove \ref{concavity_MI}, we use the fact that the mutual information depends on $\px$ only through $\py$ and that the additional term, which enforces the normalization of the probability, is only linear in $\px$, that is,
    \begin{equation}
        \func{0} = - \int dy\,\py\pa{y}\ln\pa{\py\pa{y}} + F\pasq{\px},
    \end{equation}
    where $F$ is linear in $\px$.
    \begin{IEEEproof}
        The Hessian of $\func{0}$ reads
        \begin{align}
            &\frac{\delta^{2}}{\delta \px\pa{x}\delta \px\pa{x^{\prime}}} \func{0}\pasq{\px,\nu}\\ = &- \frac{\delta^{2}}{\delta \px \pa{x}\delta \px \pa{x^{\prime}}} \int dy \left[\left(\int dx\,\pyx\left(y - x\right) \px\left(x\right)\right)\right.\nonumber\\ &\times\left.\ln\left(\int dx\,\pyx\left(y - x\right)\px\left(x\right)\right)\right]\\
            =& -\frac{\delta}{\delta \px\left(x^{\prime}\right)}\Bigg[\underbrace{\int dy\,\pyx\left(y - x\right)}_{=1}\nonumber\\+&\int dy\,\pyx\left(y - x\right)\ln\left(\int dx^{\prime}\,\pyx\left(y - x^{\prime}\right)\px\left(x^{\prime}\right)\right)\Bigg]\\
            =&- \int dy\,\frac{\pyx\left(y - x\right)\pyx\left(y - x^{\prime}\right)}{\int dx\,\pyx\left(y - x\right)\px\left(x\right)}.
        \end{align}
        Taking some test function $\varphi\pa{x}$ and sandwiching the Hessian with it, we obtain
        \begin{align}
            &-\int dx\, \varphi\pa{x}\,\int dx^{\prime}\,\varphi\pa{x^{\prime}}\int dy\,\frac{\pyx\left(y - x\right)\pyx\left(y - x^{\prime}\right)}{\int dx\,\pyx\left(y - x\right)\px\left(x\right)}\\
        =	&-\int dy\,\frac{\left(\int dx\,\varphi\left(x\right)\pyx\left(y - x\right)\right)^{2}}{\py\pa{y}}\leq0.
        \end{align}
        In other words, $\func{0}$ is concave in $\px$. Furthermore, the above inequality is fulfilled with equality if and only if
        \begin{align}
            &\int dx\, \varphi\pa{x}\pyx\left(y -x\right)	=0\ \ \forall y\in\left[-b,1+b\right]\\
\Leftrightarrow &\int_{\max\left\{ y-b,0\right\} }^{\min\left\{ y+b,1\right\} }dx\, \varphi\pa{x}	=0\ \ \forall y\in\left[-b,1+b\right].\label{eq:variation_vanishes_semilocally}    
        \end{align}
        We will demonstrate by contradiction that $\varphi\pa{x}=0$ almost everywhere, first for all $x\in \pasq{0,2b}$. Assume w.l.o.g. that 
        \begin{equation}
            \exists \, A\subseteq\left[0,2b\right] \; \mathrm{with} \;\mu\pa{A}>0 \; \mathrm{s.t.}\;\varphi\pa{x}\overset{\mathrm{assumption}}{>}0,\label{eq:assumption_variation_does_not_vanish}  
        \end{equation}
        where $\mu\pa{A}$ denotes the Lebesgue measure of $A$.
        Due to $\mu\pa{A}>0$ and the definition of a measurable set, we can assume w.l.o.g. that A is an interval, say $A=\left[x_{0},x_{0}+\epsilon\right]$, $x_{0}+\epsilon\leq 2b$. Choosing $y=x_{0}-b$, we have due to \eqref{eq:variation_vanishes_semilocally} that
        \begin{equation}
            \int_{\max\{y-b,0\}}^{\min\{y+b,1\}}dx \, \varphi(x) = \int_{0}^{x_{0}}dx\,\varphi(x) 
            =0,    
        \end{equation}
        but, again due to \eqref{eq:variation_vanishes_semilocally}, also
        \begin{equation}
            \int_{\max\{y+\epsilon-b,0\}}^{\min\{y+\epsilon +b,1\}}dx \, \varphi(x)
            =\int_{0}^{x_{0}+\epsilon}dx\, \varphi\pa{x} 
            =0,
        \end{equation}
        and therefore
        \begin{equation}
            \int_{x_{0}}^{x_{0}+\epsilon}dx\,\varphi(x)=0.
        \end{equation}
        This is in contradiction to the assumption \eqref{eq:assumption_variation_does_not_vanish}.
        We now proceed to demonstrate in the same way that $\varphi\pa{x}$ vanishes almost everywhere on the set $\pasq{0,4b}$, then on $\pasq{0,6b}$ and so on, until we cover $\left[0,1\right]$.
        Finally, we observe that the difference between $\func{}$ and $\func{0}$ is linear in $\px$, therefore also $\func{}$ is strictly concave in $\px$.
        This proves part \ref{concavity_MI} of the lemma.
        
        Note that this reasoning is specific to our case.
        If we close the interval $\pasq{0,1}$ to a circle by choosing periodic boundary conditions, then every function with mean $0$ and period $2b$ fulfills $\int dx\, \varphi\pa{x}\pyx\pa{y - x}=0 \ \forall y$.
        
        For part \ref{lbda_to_pX_injective}, we use that $\frac{\delta \func{0}}{\delta \px}$ is an injective function and therefore can be inverted (and the inverse is injective as well). Therefore, the equation $\frac{\delta \func{0}}{\delta \px\pa{x}}\overset{!}{=}\lambda\, c\pa{x}$ has a different solution for a different $\lambda$, which was to be proven.
    \end{IEEEproof}

\subsection{Proof of \Cref{lem:chain_non_overlapping} (Uniqueness of $\lambda^{\ast}$)}
\label{subsec:uniqueness_lambda_ast}
    %We use again the fact that we can find the capacity-achieving input distribution by solving the dual problem \eqref{eq:DualProblem}.
We swap the minimum and the supremum in \eqref{eq:px_max_inf_problem} and consider what is known as the dual problem
\begin{equation}
    \underset{\nu, \, \lambda \geq 0}{\inf} \, \underset{\px\geq 0}{\max}\, \func{}\pasq{\px,\nu,\lambda}.\label{eq:DualProblem}
\end{equation}
We verify that this does not change the solution by showing that Slater's condition is satisfied.
\begin{IEEEproof}
    Concretely, we check that the unconstrained optimization problem is indeed strictly concave, which is guaranteed by \lemref{convexity_px} \ref{concavity_MI}, and that by choosing $\px(x) = \delta\left(x\right)$, we indeed strictly fulfill any cost constraint with $\bar{c}>0$. 
    %Indeed, according to \lemref{concavity_MI}, the unconstrained optimization problem is strictly concave and, by choosing $\px = \delta\left(x\right)$, we strictly fulfill any cost constraint with $\bar{c}>0$. Together, this means that Slater's condition is fulfilled and 
    Therefore, Slater's theorem applies and yields that the dual problem has the same solution as the original one (cf. chap. 5.2.3 in \cite{Boyd_convex_optimization_2004}).
    We can hence assume that we optimize $\px$ for a fixed $\lambda$.\\
    Now, observing that the respective last terms of \eqref{eq:def_constr_func} and \eqref{eq:def_unconstr_func} do not depend on $\px$, we define
    \begin{align}
        \tilde{\func{}}\pasq{\lambda,\nu} \coloneqq &\underset{\px\geq 0}{\max} \func{}\pasq{\px,\nu,\lambda} - \nu  - \lambda \cbar,
    \end{align}
    where $\func{}$ is given by \Cref{Def_Ls}.
    Considering the convex combination of $\tilde{\func{}}$ evaluated at two different points, we have for $t\in\left[0,1\right]$
    \begin{align}
        &t \tilde{\func{}}\pa{\nu_{0},\lambda_{0}} + \left(1-t\right) \tilde{\func{}}\pa{\nu_{1},\lambda_{1}}\\ 
        = &t\,  \underset{\px\geq 0}{\max}\, \func{}\pasq{\px,\nu_{0},\lambda_{0}} + \left(1-t\right) \underset{\px\geq 0}{\max}\,\func{}\pasq{\px,\nu_{1},\lambda_{1}}\\
        \geq & t\,  \func{}\pasq{\tilde{p}_{X},\nu_{0},\lambda_{0}} + \left(1-t\right)\func{}\pasq{\tilde{p}_{X},\nu_{1},\lambda_{1}}\label{eq:proof_unique_lbda_ineq_to_take_same_px}
    \end{align}
    for an arbitrary distribution $\tilde{p}_{X}$. This inequality is still fulfilled after building the maximum over $\tilde{p}_{X}$, so that we obtain, by regrouping the terms
    \begin{align}
        &t \tilde{\func{}}\pa{\nu_{0},\lambda_{0}} + \left(1-t\right) \tilde{\func{}}\pa{\nu_{1},\lambda_{1}}\\
        \geq&\underset{\tilde{p}_{X}\geq 0}{\max} \pacu{t\,  \func{}\pasq{\tilde{p}_{X},\nu_{0},\lambda_{0}} + \left(1-t\right)\func{}\pasq{\tilde{p}_{X},\nu_{1},\lambda_{1}}}
        \\
        = &\underset{\tilde{p}_{X}\geq 0}{\max} \,\left\{  t \left[I\pasq{\tilde{p}_{X}} - \nu_{0} \int dx\, \tilde{p}_{X}\pa{x}\right.\right.\nonumber\\
        & \hspace{2.35cm} \left. -{} \lambda_{0} \int dx\, \tilde{p}_{X}\pa{x}c\pa{x} \right]\\ 
        + &\pa{1-t}\, \left[I\pasq{\tilde{p}_{X}} - \nu_{1} \int dx\, \tilde{p}_{X}\pa{x}\right.\nonumber\\
        & \hspace{2.35cm} \left.\left. -{}\lambda_{1} \int dx\, \tilde{p}_{X}\pa{x}c\pa{x} \right] \right\}\\
        =&\tilde{\func{}}\pa{t\nu_{0} + \pa{1-t}\nu_{1},t\lambda_{0}+\pa{1-t}\lambda_{1}}.
    \end{align}
    In words: $\tilde{\func{}}$ is convex. Furthermore, according to \lemref{convexity_px}, \ref{lbda_to_pX_injective}, different $\lambda$ result in different maximizing $\px$. Therefore, the inequality in \eqref{eq:proof_unique_lbda_ineq_to_take_same_px} is sharp and thus $\tilde{\func{}}$ even strictly convex. Therefore, the condition on $\pa{\nu, \lambda}^\mathrm{T}$ resulting from the supremum in \eqref{eq:DualProblem},
    \begin{eqnarray}
        \frac{d \pa{\underset{\px\geq 0}{\max}\func{} \pasq{\px, \nu, \lambda}}}{d \pa{\nu, \lambda}^\mathrm{T}} \overset{!}{=} & 0\\     
        \Leftrightarrow\frac{d \tilde{\func{}} \left(\pa{\nu, \lambda}^\mathrm{T}\right)}{d \pa{\nu, \lambda}^\mathrm{T}} \overset{!}{=} &-\pa{\begin{matrix} 1\\ \cbar\end{matrix}}\label{eq:Defining_equation_lbda_to_cbar}
    \end{eqnarray}
    has a unique solution. This completes the proof.
    \end{IEEEproof}
    In fact, in the same way as we have demonstrated that $\tilde{\func{}}$ is strictly convex in $\pa{\nu, \lambda}^\mathrm{T}$, we can show that $\tilde{\func{}}$ is also convex in $\nu$ for a fixed $\lambda$ and therefore we find a unique $\nu\pa{\lambda}$ for all valid $\lambda$. Consequently, we can write $\tilde{\func{}}$ as $\tilde{\func{}}\pa{\nu\pa{\lambda},\lambda}$ and
    \begin{equation}
        \frac{d\tilde{\func{}}}{d\lambda} \overset{!}{=} -\cbar - \frac{\partial \nu}{\partial \lambda},\label{eq:def_fct_lbda_cbar}
    \end{equation}
    which defines an injective function $\lambda^{\ast} \rightarrow \cbar$ whose derivative is
    \begin{equation}
        \frac{d^2}{d\lambda^{2}}\tilde{\func{}}\left(\nu\left(\lambda\right),\lambda\right) = \underbrace{\left(\begin{matrix} \frac{d\nu}{d\lambda}\\ 1
        \end{matrix}\right)^{\mathrm{T}} \mathrm{H}\tilde{\func{}} \left(\begin{matrix} \frac{d\nu}{d\lambda}\\ 1
        \end{matrix}\right)}_{> 0} + \underbrace{\pa{\frac{\partial \tilde{\func{}}}{\partial \nu} - 1}}_{=0} \frac{d^{2}\nu}{d\lambda^{2}},
    \end{equation}
    where $H\tilde{\func{}}$ denotes the Hessian of $\tilde{\func{}}$. The last term vanishes because we evaluate $\tilde{\func{}}$ at its extremizing arguments.
    Taking this together with \eqref{eq:def_fct_lbda_cbar}, we conclude that $\cbar\pa{\lambda^\ast}$ is a strictly decreasing function. 
    
    In the integer case IIa, furthermore, we can even show this by explicitly differentiating $\cbar\pa{\lambda}=\Ex{c\pa{x}}_{\px^\lambda}$ with respect to $\lambda$.
    %%%%%%%%%%%%%%%%%%%% proof uniqueness of lambda r in N %%%%%%%%%%%%%%%%%%
%\subsection{Explicit Proof of \lemref{chain_non_overlapping} (uniqueness of $\lambda^{\ast}$ for $r\in \mathbb{N}$)}
\begin{IEEEproof}
By construction, for a given $\lambda$, the $N_r$ masses
$m_{j}$ fulfill the $N_r-1$ difference equations (\ref{eq:diff_eqs_integer_case})
and the corresponding cost is given by $\left\langle c\left(x\right)\right\rangle _{\px^{\lambda}}$.
Combined with the cost constraint $\bar{c}=\Ex{c\pa{x}}_{\px^{\lambda}}$, this is equivalent to the $N_r$ original equations \eqref{eq:eq_constr_integer_case}. 
When $\lambda=0$, the constraint is inactive and $m_{j}\left(0\right)=1/\left(n+1\right)$,
and $\left\langle c\left(x\right)\right\rangle _{\px^{0\ast}}=\bar{c}^{\ast}$.
In the opposite limit of $\lambda\rightarrow\infty$, all the probability
is concentrated at zero, i.e. $\lim_{\lambda\rightarrow\infty}m_{1}\left(\lambda\right)=1$,
and for all other masses, $j>1$, $\lim_{\lambda\rightarrow\infty}m_{j}\left(\lambda\right)=0$,
which yields $\lim_{\lambda\rightarrow\infty}\Ex{c\pa{x}}_{\px^{\lambda}}=0$.
In between the two extremes, $\Ex{c\pa{x}}_{\px^{\lambda}}$ is a strictly monotonic decreasing function.
We obtain
\begin{align}
&\frac{\partial}{\partial\lambda}\left\langle c\left(x\right)\right\rangle _{\px^{\lambda}} =\frac{\partial}{\partial\lambda}\sum_{j=1}^{N_r}m_{j}\left(\lambda\right)\,c_{j}
 =\frac{\partial}{\partial\lambda}\sum_{j=1}^{N_r} c_{j}\frac{e^{-\lambda c_{j}}}{z\left(\lambda\right)}\nonumber\\
 =&\sum_{j=1}^{N_r} c_{j}\frac{-c_{j}e^{-\lambda c_{j}}z\left(\lambda\right)+e^{-\lambda c_{j}}\sum_{l}c_{l}e^{-\lambda c_{l}}}{z^{2}\left(\lambda\right)}\nonumber\\
% =&-\sum_{j} c_{j}^{2} \, m_{j}\left(\lambda\right)
% +\biggl(\sum_{j}m_{j}\left(\lambda\right)\,c_{j}\biggr)\biggl(\sum_{k}m_{k}\left(\lambda\right)\,c_{k}\biggr)\nonumber\\
 =&-\left(\left\langle c^{2}\left(x\right)\right\rangle _{\px^{\lambda}}-\left\langle c\left(x\right)\right\rangle _{\px^{\lambda}}^{2}\right)
 =-\mathrm{Var}_{\px^{\lambda}}\left(c\right)\leq0,
\end{align}
with equality if and only if the total mass is concentrated on $m_{1}$,
i.e. in the case $\lambda\rightarrow\infty$. Therefore, if $\bar{c}\in(0,\bar{c}^\ast]$ there is one unique
$\px^{\lambda^{\ast}}$ such that $\left\langle c\left(x\right)\right\rangle _{\px^{\lambda^{\ast}}}=\bar{c}$.
\end{IEEEproof}
%%%%%%%%%%%%%%%%%%%%%%%%%%%%%%%%%%%%%%%%%%%%%%%%%%%%%%%%%%%%%%%%%%%%%%%%%

%%%%%%%%%%%%%%%%%%%% proof piecewise linearity of i(x) %%%%%%%%%%%%%%%%%%
\section{Proof of \lemref{linear_i} (piecewise linearity of the marginal information density)}
\label{subsec:proof_i_linear}
Here, we will prove the piecewise linearity of the marginal information density.

\textit{Let the positions $x_{j}$ be as defined in \eqref{eq:def_pos_unconstr}, and the corresponding masses $m_{j}\geq0$.
Then $i\pa{x;\px}$ is linear for $x\in\left[x_{j},x_{j+1}\right]$ and the slopes of the linear segments are given by
\begin{enumerate}
    \item $r \log\pa{\frac{m_j}{m_{j+1}}}$ if $r\in \mathbb{N}$;\label{itm:lem_lin_i_1}
    \item $r\log\left[\left(m_{j-1}+m_{j}\right)/\left(m_{j+1}+m_{j+2}\right)\right]$ if $r\notin \mathbb{N}$,\label{itm:lem_lin_i_2}
\end{enumerate}
and if the respective denominators are positive. Here, $j\in\rangeone{N_r-1}$ for \ref{itm:lem_lin_i_1} and $j\in\rangeone{N_r-2}$ for \ref{itm:lem_lin_i_2}.}
\begin{IEEEproof}
    We evaluate the marginal information density as defined in \eqref{eq:def_info_density} for $x\in \pasq{x_j,x_{j+1}}$ and use $d\coloneqq x-x_{j}$.
\begin{enumerate}
    \item For $r\in \mathbb{N}$ the output distribution $\py\pa{y;\px}$ consists of two constant segments from $x-b$ to $x_{j}+b$ and from $x_{j}+b$ to $x+b$ (cf. Figure \ref{fig:sketch_integer_case} Ia).
    We obtain
    \begin{align}
        i\pa{x;\px}  =&-\frac{1}{2b}\int_{x-b}^{x+b}dy\,\log\pasq{2b\,\py\pa{y;\px}}\\
        =&-\frac{1}{2b}\int_{x-b}^{x_j+b}dy\,\log m_j\nonumber\\
        & \;{}-\frac{1}{2b}\int_{x_j+b}^{x+b}dy\,\log m_{j+1}\\
        =&r \log\pa{\frac{m_j}{m_{j+1}}}\,d - \log m_j,
    \end{align}
    where we used $r=1/\pa{2b}$.
    \item Consider the case $r\notin \mathbb{N}$.
    For $x\in\left[x_{j},x_{j+1}\right]$, $\py\pa{y;\px}$ consists
    of three piece-wise constant segments between the positions $x-b \leq x_{j+1}-b\leq x_j+b \leq x+b$ (cf. Figure \ref{fig:sketch_integer_case} Ib).
    The marginal information density evaluates to
\begin{align}
i\left(x;\px\right) & =-\frac{1}{2b}\int_{x-b}^{x+b}dy\,\log\left[2b\,\py\pa{y;\px}\right]\\
 & =-r\left(x_{j+1}-x_j -d\right)\log\left(m_{j-1}+m_{j}\right)\nonumber\\
 & \hphantom{=}-r\left(2b-x_{j+1}+x_j\right)\log\left(m_{j}+m_{j+1}\right)\nonumber\\
 & \hphantom{=}-r\,d\log\left(m_{j+1}+m_{j+2}\right)\nonumber\\
 & =r\log\left(\frac{m_{j-1}+m_{j}}{m_{j+1}+m_{j+2}}\right)d+D,
\end{align}
where all terms independent of $d$ are absorbed into $D$.
\end{enumerate}
\end{IEEEproof}
%%%%%%%%%%%%%%%%%%%%%%%%%%%%%%%%%%%%%%%%%%%%%%%%%%%%%%%%%%%%%%%%%%%%%%%%%

\section{Proof of Case II$\mathrm{b}$}
\label{sec:proof_case_IIb}
\subsection{Prerequisites}
\label{appendix_A_prerequisites}
To simplify the proof of Case IIb, we first derive general properties of the exponents $\dhat{}$ and $\dbar{}$ and of the probability masses $m_j$. We start with the definition of the exponents $\dhat{}$ and $\dbar{}$, which occur in the expression in \eqref{eq:m_hat_j_expl_general} and \eqref{eq:m_bar_j_expl_general} of the combined masses $\mhat{j}$ and $\mbar{j}$.
%%%%%%%%%%%%%%%%%%%%%%%%% definition dhat dbar %%%%%%%%%%%%%%%%%%%%%%%%%%
\begin{definition}
    For fixed $k\in\range{0}{n-1}$ define
    \begin{align}
    \dhat{j}^{k} &\coloneqq \begin{cases}
    0, & j =  k+1\\
        \frac{1}{r}\sum_{i=k+1}^{j-1} \frac{c_{2i} - c_{2i-1}}{x_{2i}-x_{2i-1}}, & j>k+1,
    \end{cases} \label{eq:definition_dhat_general}\\
    \dbar{\ell}^{k} &\coloneqq \begin{cases}
    0, & \ell =  k+1\\
        \frac{1}{r}\sum_{i=k+1}^{j-1} \frac{c_{2i+1} - c_{2i}}{x_{2i+1}-x_{2i}}, & \ell>k+1,
    \end{cases} \label{eq:definition_dbar_general}
\end{align}
where $c_j\coloneqq c\pa{x_j}$. Indices $j<k+1$ are not needed since for fixed $k$, the summations in $\mhat{}$ and $\mbar{}$ depend only on $\dhat{j}$ and $\dbar{j}$, $j\geq k+1$.
\end{definition}
These definitions of $\dhat{j}^{k}$ and $\dbar{j}^{k}$ depend on $k$ but we will drop the index $k$ whenever clear from context.
%Using $\rho \coloneqq r - n + 1$, the differences between the positions in the denominator can also be written as $x_{2i} - x_{2i-1}=\rho/r$ and $x_{2i+1} - x_{2i} = \pa{1-\rho}/r$ independent of the summation index.
%Inserting this into \eqref{eq:definition_dhat_general} and \eqref{eq:definition_dbar_general} allows us to write $\mhat{}$ and $\mbar{}$ as shown in \thmref{main}, i.e. eqs.~\eqref{eq:m_hat_j_expl_general} and \eqref{eq:m_hat_j_expl_general}.
The case $\cbar \in (\theta_0,\cbar^\ast]$, where all masses are present, is retrieved for $k=0$.
%%%%%%%%%%%%%%%%%%%%% remark Boltzmann distribution %%%%%%%%%%%%%%%%%%%%%
\begin{remark}
\label{rem:boltzmann_interpretation}
    We can interpret slightly modified versions of the combined masses as probabilities of Boltzmann distributions. To see this, define
    \begin{align}
\mhat{j}^{\prime} & \coloneqq\frac{1}{\zhat^{\prime}}e^{-\lambda^\ast \dhat{j}}, & \zhat^{\prime}\coloneqq & \sum_{j=k+1}^{n+1}e^{-\lambda^\ast \dhat{j}},\\
\mbar{\ell}^{\prime} & \coloneqq \frac{1}{\zbar^{\prime}}e^{-\lambda^\ast \dbar{\ell}}, & \zbar^{\prime} \coloneqq & \sum_{j=k+1}^{n}e^{-\lambda^\ast \dbar{j}},\label{eq:m_bar_prime}
\end{align}
    so that $\mhat{j}^{\prime}=\mhat{j}/\mkg{k}$ and $\mhat{\ell}^{\prime}=\mhat{\ell}/\mkg{k}$ for $j\in\range{k+1}{n+1}$ and $\ell\in\range{k+1}{n}$.
    Since we excluded the normalization factor $\mkl{k}$ in the definition in these definitions, $\mhat{j}^{\prime}$ and $\mbar{j}^{\prime}$ sum to one, are positive and follow an exponential form with the exponents $\dhat{j}$ and $\dbar{j}$.
    Therefore, they can be interpreted as Boltzmann weights for two systems $\hat{U}$ and $\bar{U}$ with energy levels $\dhat{}\coloneqq (\dhat{k+1},\,\ldots,\,\dhat{n+1})$ and $\dbar{}\coloneqq (\dbar{k+1},\,\ldots,\,\dbar{n})$, respectively.
    The energies are cumulative sums of the differences between the cost function evaluated at neighboring $x_j$ and the Lagrange multiplier $\lambda$ plays the role of the inverse temperature.
    Similarly, the rescaled versions of masses $m_j\in\Mkl{k}$ can be defined as
    \begin{align}
        m_{2j-1}^{\prime} & \coloneqq \frac{1}{z^{\prime}}e^{-\lambda^{\ast}c_{2j-1}}, & z^{\prime} \coloneqq & \sum_{j=1}^{k} e^{-\lambda^{\ast}c_{2j-1}},
    \end{align}
    where $j\in\rangeone{k}$. They also can be interpreted as Boltzmann weights of a distribution that we will refer to as $U$.
\end{remark}
%%%%%%%%%%%%%%%%%%%%%%% proof m_2 vanishes first %%%%%%%%%%%%%%%%%%%%%%%%
\label{subsec:m_2_vanishes_first}
To prove that the position $x_2$ is the first to drop from the support as the cost budget $\cbar$ tightens, we need two auxiliary lemmas.
They collect basic properties of the exponents $\dhat{}$ and $\dbar{}$ and, in turn, of the combined masses $\mhat{}$ and $\mbar{}$.
%%%%%%%%%%%%%%%%%%%%%%% lemma properties dhat dbar %%%%%%%%%%%%%%%%%%%%%%%
\begin{lemma}[Properties of $\dhat{}$ and $\dbar{}$]
    \label{lem:properties_ds}Let $c\pa{x}$ be a convex cost function as specified in \Cref{ass:cost_function} ($\alpha\leq1$) and let $\dhat{j}$ and $\dbar{j}$ as defined in \eqref{eq:definition_dhat_general} and \eqref{eq:definition_dbar_general}. With $0<\rho<1$, it holds that
    \begin{enumerate}
        \item $\dhat{j} \leq \dhat{j+1}$ and $\dbar{\ell} \leq \dbar{\ell+1}$ for $j\in \range{k+1}{n}$ and $\ell\in \range{k+1}{n-1}$;
        \item $\dbar{j} \leq \dhat{j}$ for $j\in \range{k+1}{n}$, with equality for $j>k+1$ if and only if $\alpha=1$.
    \end{enumerate}
\end{lemma}
\begin{IEEEproof}
For $j\leq k+1$, where $\dhat{j}=\dbar{j}=0$, both assertions are trivially satisfied.
In the remainder of this proof we assume $j>k+1$, which requires a more detailed analysis. 
\mbox{}\par  % force a line break after the "Proof." heading
\begin{enumerate}[label=\textit{\arabic*)},leftmargin=*,itemsep=0.25em,topsep=0.25em]
    \item Due to the cumulative sums in the definitions of $\dhat{}$ and $\dbar{}$ we obtain
        \begin{equation}
            \dhat{j+1} - \dhat{j} = \frac{1}{\rho} \pasq{c\pa{x_{2j}} - c\pa{x_{2j-1}}}> 0,
        \end{equation}
        and
        \begin{equation}
            \dbar{j+1} - \dbar{j} = \frac{1}{1-\rho} \pasq{c\pa{x_{2j+1}} - c\pa{x_{2j}}}> 0,
        \end{equation}
        because $c\pa{x}$ is an increasing function of $x$.
    \item The different terms in the definitions \eqref{eq:definition_dhat_general} and \eqref{eq:definition_dbar_general} for $\dhat{}$ and $\dbar{}$ can be written in terms of integrals, which yields
    \begin{align}
        \dhat{j} &\coloneqq
        \frac{1}{r}\sum_{i=k+1}^{j-1} \underbrace{\frac{1}{x_{2i} - x_{2i-1}}\int_{x_{2i-1}}^{x_{2i}} c^\prime\pa{x}\,dx}_{T_{1}^{\pa{i}}},
        \label{eq:definition_dhat_general_integ}\\
        \dbar{j} &\coloneqq 
        \frac{1}{r}\sum_{i=k+1}^{j-1} \underbrace{\frac{1}{x_{2i+1} - x_{2i}}\int_{x_{2i}}^{x_{2i+1}}c^\prime\pa{x}\,dx}_{T_{2}^{\pa{i}}}.    \label{eq:definition_dbar_general_integ}
\end{align}

To prove that $\dhat{j}\geq\dbar{j}$, it is sufficient to show that the inequality holds termwise for every $i$.
The integrals $T_{1}^{\pa{i}}$ and $T_{2}^{\pa{i}}$ are averages of the derivative of the cost function over the two consecutive intervals.
By the mean value theorem, there exist $\xi^{\pa{i}} \in \pasq{x_{2i-1},\, x_{2i}}$ and $\eta^{\pa{i}} \in \pasq{x_{2i},\, x_{2i+1}}$ so that $T_{1}^{\pa{i}}=c^{\prime}\pa{\xi^{\pa{i}}}$ and $T_{2}^{\pa{i}}=c^{\prime}\pa{\eta^{\pa{i}}}$.
Since $\xi^{\pa{i}}>\eta^{\pa{i}}$, and $c^{\prime}$ is a non-increasing function due to the concavity of $c\pa{x}$, we conclude that $T_{1}^{\pa{i}}\geq T_{2}^{\pa{i}}$ and therefore $\dhat{j}\geq\dbar{j}$ for $0<\alpha\leq 1$ and $0<\rho<1$. Equality can only occur if $c\pa{x}$ is linear ($\alpha=1$) because it is assumed to be continuous and differentiable. In the linear case, we obtain
\begin{equation}
    \dhat{j}=\dbar{j}=\frac{j-k-1}{r},
\end{equation}
for $j\in \range{k+2}{n}$, and additionally $\dhat{n+1}=\frac{n-k}{r}$.

\end{enumerate}
\end{IEEEproof}
%%%%%%%%%%%%%%%%%%%%%%%%%%%%%%%%%%%%%%%%%%%%%%%%%%%%%%%%%%%%%%%%%%%%%%%%%
From these properties of the exponents, we can derive some useful relations of the combined masses $\mhat{}$ and $\mbar{}$, and of the masses $m_j$.
%%%%%%%%%%%%%%%%%%%%%%% lemma properties mhat mbar %%%%%%%%%%%%%%%%%%%%%%%
\begin{lemma}[Properties of the masses]
\label{lem:properties_masses}
Let $c\pa{x}$ be a strictly concave cost function as specified in \Cref{ass:cost_function} ($\alpha<1$) and let $r\notin\mathbb{N}$, and $\bar c\in(\theta_0,\bar c^{\ast}]$.
Let the masses $m_j(\lambda)\in\Mkg{k}$, be defined by \eqref{eq:def-masses-non-integer-case}, where $\mhat{}$ and $\mbar{}$ and the corresponding normalizations $\zhat$ and $\zbar$ are given by \eqref{eq:m_hat_j_expl_general} and \eqref{eq:m_bar_j_expl_general}.
If $\zhat > \zbar$, then the combined masses satisfy
\begin{enumerate}[label=\arabic*)]
  \item
  \begin{equation}
      \mbar{j}\geq\mhat{j}\geq\mhat{n+1},
  \end{equation}
  for $j\in \range{k+2}{n}$;
  \item the odd-indexes masses are decreasing and the even-indexed masses are increasing with increasing index, i.e.
  \begin{align}
    m_{j} &> m_{2n-1}, & 2k+1\leq j \leq 2n-3, & & j\;\mathrm{odd},\\
    m_{j} &> m_{2k+2}, & 2k+4\leq j \leq 2n, \hphantom{-3}& & j\;\mathrm{even},
  \end{align}
  see also \figref{sketch_integer_case} Ib / IIb.
\end{enumerate}
\end{lemma}
\begin{IEEEproof}
%\mbox{}\par  % force a line break after the "Proof." heading
\begin{enumerate}[label=\textit{\arabic*)},leftmargin=*,itemsep=0.25em,topsep=0.25em]
    \item Using the assumption $\zhat>\zbar$ and \lemref{properties_ds}, we have
    \begin{equation}
        \mbar{j} = \frac{e^{-\lambda \dbar{j}}}{\zbar}>\frac{e^{-\lambda \dbar{j}}}{\zhat} \geq \frac{e^{-\lambda \dhat{j}}}{\zhat}=\mhat{j}.
    \end{equation}
    Since $\mhat{n+1}/\mhat{l}=e^{-\lambda\pa{\dhat{n+1}-\dhat{k}}}\leq 1$ by \lemref{properties_ds}, we see that $\mhat{n+1}$ is a lower bound for all combined masses.
    \item Using the above result and the back transform \eqref{eq:def-masses-non-integer-case} to the original masses yields
    \begin{equation}
        m_{2j+1} - m_{2j-1} = \mhat{j+1} - \mbar{j} <\mhat{j} - \mbar{j}<0,
    \end{equation}
    for $j\in \range{k+1}{n}$. Similarly,
    \begin{equation}
        m_{2j+2} - m_{2j} = \mbar{j+1} - \mhat{j+1} > 0,
    \end{equation}
    for $l\in \range{k+1}{n-1}$.
\end{enumerate}
\end{IEEEproof}
\begin{corollary}
\label{cor:smallest_masses}
The mass $m_{2n-1}$ the smallest of the odd-indexed masses and $m_{2k+2}$ is the smallest of the even-indexed masses in $\Mkg{k}$.
\end{corollary}
\begin{IEEEproof}
    This follows directly since the odd-indexed masses decrease with the index and the even-indexed increase.
\end{IEEEproof}
%%%%%%%%%%%%%%%%%%%%%%%%%%%%%%%%%%%%%%%%%%%%%%%%%%%%%%%%%%%%%%%%%%%%%%%%%
Finally, we consider a property of the normalization factors $\mkl{k}$ and $\mkg{k}$.
We expect $\mkl{k}$ to increase and $\mkg{k}$ to decrease with tightening cost budget $\cbar$ because probability mass is shifted to the lower cost positions of $\xkl{k}$.
%%%%%%%%%%%%%%%%%%%%%%% properties of m_< / m_> %%%%%%%%%%%%%%%%%%%%%%%%%
\begin{lemma}[Strict monotonic increase of $\mkl{k}/\mkg{k}$]
\label{lem:properties_mkl_mkg}
    Let $c\pa{x}$ be a convex cost function as specified in \Cref{ass:cost_function} ($\alpha\leq1$), $r\notin\mathbb{N}$ and $\bar{c}\in(\theta_{k},\theta_{k-1}]$. Let the masses $m_{j}\left(\lambda\right)$ and the normalization factors $\mkl{k}\pa{\lambda}$ and $\mkg{k}\pa{\lambda}$ given by \eqref{eq:m_smaller}-\eqref{eq:m_bar_j_expl_general}. Then
    \begin{equation}
        \frac{\partial}{\partial\lambda}\frac{\mkl{k}\pa{\lambda}}{\mkg{k}\pa{\lambda}} > 0 \:\, \mathrm{\textit{and}}\:\,\frac{\partial}{\partial \lambda}\mkl{k}\pa{\lambda}>0, \label{eq:normalization_ratio_derivative}
    \end{equation}
    for $\lambda \geq 0$ and $k\in\rangeone{n-2}$.
\end{lemma}
\begin{IEEEproof}
    We determine the ratio $\mkl{k}/\mkg{k}$ by taking the difference between the equality constraint \eqref{eq:eq_constr} evaluated at $x_1$ and at $x_{2k+1}$. Using the parametrization \eqref{eq:general_ansatz} of $\px$, the marginal information density \eqref{eq:def_info_density} at these two points can be written as
    \begin{IEEEeqnarray}{rCl}
        i\pa{x_1} &=& -\log\pasq{m_1}\\
        i\pa{x_{2k+1}} &=& -\rho\log\pasq{\mhat{k+1}}  {}-\pa{1-\rho}\log\pasq{\mbar{k+1}}.
    \end{IEEEeqnarray}
    Together with the r.h.s. of the equality constraint, i.e. $I+\lambda\pa{c\pa{x}-\cbar}$, we obtain
    \begin{equation}
        i\pa{x_{2k+1}}-i\pa{x_{1}}=\lambda\pa{c_{2k+1}-c_1}.\label{eq:i_diff}
    \end{equation}
    In the following, instead of considering $z$, $\zbar$, and $\zhat$, we use the rescaled versions $z^{\prime}$, $\zhat^{\prime}$, and $\zbar^{\prime}$ as introduced in \Cref{rem:boltzmann_interpretation}.
    % \begin{IEEEeqnarray}{rCl}
    %     u &=&\sum_{j=1}^{k}e^{-\lambda^\ast c_{2j-1}} = z \cdot \mkl{k}, \\
    %     \hat u & = & \sum_{j=k+1}^{n+1}e^{-\lambda^\ast \dhat{j}} = \zhat \cdot \mkg{k}, \\
    %     \bar u &=& \sum_{j=k+1}^{n}e^{-\lambda^\ast \dbar{j}} = \zbar \cdot \mkg{k},
    % \end{IEEEeqnarray}
    We obtain $m_1 = z^{-1} = \mkl{k}/z^{\prime}$, $\mhat{k+1} = \zhat^{-1} = \mkl{k}/\zhat^{\prime}$, and $\mbar{k+1}= \zbar^{-1}= \mkl{k}/\zbar^{\prime}$.
    Solving \eqref{eq:i_diff} for the ratio $\mkl{k}/\mkg{k}$ gives
    \begin{IEEEeqnarray}{rCl}
        \frac{\mkl{k}\pa{\lambda}}{\mkg{k}\pa{\lambda}} &=&\exp\big[\lambda\pa{c_{2k+1}-c_1}+\log\pa{z^{\prime}}\nonumber\\ && {}-\rho \log \pa{\zhat^{\prime}} - \pa{1-\rho} \log \pa{\zbar^{\prime}}\big].\label{eq:normalization_ratio}
    \end{IEEEeqnarray}
    To prove \eqref{eq:normalization_ratio_derivative}, we argue that the exponent in \eqref{eq:normalization_ratio} strictly increases with $\lambda>0$. The terms in the second line are increasing because
    \begin{align}
        -\frac{\partial}{\partial\lambda}\log\pa{\zhat^{\prime}}&=\underbrace{\frac{1}{\zhat^{\prime}}\sum_{j=k+1}^{n+1}\dhat{j}e^{-\lambda \dhat{j}}}_{\eqqcolon\Ex{\dhat{}}_{\mhat{}^{\prime}}}>0,\label{eq:def_expectatio_hat}
    \end{align}
    and similarly
    \begin{align}
        -\frac{\partial}{\partial\lambda}\log\pa{\zbar^{\prime}}&=\underbrace{\frac{1}{\zbar^{\prime}}\sum_{j=k+1}^{n}\dbar{j}e^{-\lambda \dbar{j}}}_{\eqqcolon\Ex{\dbar{}}_{\mbar{}^{\prime}}}>0,\label{eq:def_expectatio_bar}
    \end{align}
    where we introduced the expectation values $\Ex{\cdot}_{\mhat{}^{\prime}}$ and $\Ex{\cdot}_{\zbar^{\prime}}$ with respect to the Boltzmann distributions $\hat{U}$ and $\bar{U}$, respectively, see \Cref{rem:boltzmann_interpretation}.
    
    For the first line in \eqref{eq:normalization_ratio} we obtain
    \begin{align}
        \frac{\partial}{\partial\lambda}&\pasq{\lambda\pa{c_{2k+1}-c_1}+\log \pa{u}}\\
        &= c_{2k+1}+\frac{\partial}{\partial\lambda}\log u\\
        &=c_{2k+1}-\Ex{c}_{m^{\prime}},
    \end{align}
    where we defined the expectation value in the last line as with respect to the Boltzmann distribution $U$ introduced on \Cref{rem:boltzmann_interpretation}.
    % \begin{equation}
    %     \Ex{c}_u\coloneqq u^{-1}\sum_{j=1}^{k}c_{2j-1}\,e^{-\lambda c_{2j-1}}.
    % \end{equation}
    For a strictly increasing cost function, the (weighted) average over $\pacu{c\pa{x_j}\mid x_j \in \xkl{k}}$ is smaller than $c\pa{x_{2k+1}}$. 
    Thus, $\frac{\partial}{\partial\lambda}\pasq{\lambda\pa{c_{2k+1}-c_1}+\log\pa{u}}>0$, and therefore, the entire exponent in \eqref{eq:normalization_ratio} is an increasing function of $\lambda$.

    Since the ratio $\mkl{k}/\mkg{k}$ increases with $\lambda$ and since $\mkl{k}\pa{\lambda} + \mkg{k}\pa{\lambda} = 1$, it follows that $\partial/\partial \lambda\; \mkl{k}\pa{\lambda}>0$.
\end{IEEEproof}
%%%%%%%%%%%%%%%%%%%%%%%%%%%%%%%%%%%%%%%%%%%%%%%%%%%%%%%%%%%%%%%%%%%%%%%%%

\subsection{Proof of Case IIb}
\label{subsec:proof_case_IIb}
%%%%%%%%%%%%%%%%%%%%%%%%%%% proof of Case IIb %%%%%%%%%%%%%%%%%%%%%%%%%%%
\subsubsection{Step A}
To prove \emph{Step A}, i.e. for $\bar{c} > \theta_0$, we assume that the positions of the masses are given by (\ref{eq:def_pos_unconstr}), i.e. $S=S_0$, and show \emph{a posteriori} that those positions are optimal.
Using $\rho \coloneqq r-\floor{r}$, the differences between the positions be written as $x_{2i} - x_{2i-1}=\rho/r$ and $x_{2i+1} - x_{2i} = \pa{1-\rho}/r$ independent of the summation index.
Together with the definitions \eqref{eq:def_mhat_mbar} of the combined masses, the marginal information density defined in \eqref{eq:def_info_density} evaluates to
\begin{equation}
i\left(x_{j},\px\right) =-\rho\log\hat{m}_{\hat{f}\left(j\right)}-\left(1-\rho\right)\log\bar{m}_{\bar{f}\left(j\right)},\label{eq:i_non_integer}
\end{equation}
where the labels are given by
\begin{equation}
    \begin{split}
        \hat{f}\pa{j} &\coloneqq \floor{j/2}+1\;\mathrm{and}\\
        \bar{f}\pa{j} &\coloneqq \floor{\left(j+1\right)/2},
    \end{split}\label{eq:def_fhat_fbar}
\end{equation}
with $j\in\rangeone{N_r}$.
Inserting \eqref{eq:i_non_integer} into the equality constraint \eqref{eq:eq_constr} gives $N_r$ equations of the form
\begin{equation}
    -\rho\log\hat{m}_{\hat{f}\left(j\right)}-\left(1-\rho\right)\log\bar{m}_{\bar{f}\left(j\right)} = I+\lambda \pa{c_i-\cbar},\label{eq:eq_costr_IIb}
\end{equation}
which, together with the normalization condition $\sum_{j=1}^{N_r}m_j=1$, determine the $N_r$ masses $m_j$ and the value of the Lagrange multiplier $\lambda$.
However, this set of equations is coupled because the mutual information $I$ depends on the value of all masses.
To obtain explicit formulas for the masses, we therefore subtract the $\left(2j\right)$-th equality of from the $\left(2j-1\right)$-th equality, and subtract the $\left(2j+1\right)$-th equality of \eqref{eq:eq_costr_IIb} from the $\left(2j\right)$-th equality.
This yields 
\begin{equation}
\begin{split}
    \hat{m}_{j+1} & =\hat{m}_{j}e^{-\lambda\frac{\widehat{\Delta c}_{j}}{\rho}},\;j=1,\dots,n,\\
    \bar{m}_{j+1} & =\bar{m}_{j}e^{-\lambda\frac{\overline{\Delta c}_{j}}{1-\rho}},\;j=1,\dots,n-1,
\end{split}
\label{eq:result_diff_eq}
\end{equation}
respectively. Here, we defined 
\begin{equation}
    \begin{split}
        \widehat{\Delta c} & \coloneqq\left(c_{2}-c_{1},c_{4}-c_{3},\ldots,c_{2n}-c_{2n-1}\right),\\
        \overline{\Delta c} & \coloneqq\left(c_{3}-c_{2},c_{5}-c_{4},\ldots,c_{2n-1}-c_{2n-2}\right).
    \end{split}\label{eq:def_chat_cbar}
\end{equation}
Including the normalization \eqref{eq:sum_constraint} and using the definitions \eqref{eq:definition_dhat_general} and \eqref{eq:definition_dbar_general} of $\dhat{_j}$ and $\dbar{j}$ in Appendix \ref{appendix_A_prerequisites}, we obtain the expressions \eqref{eq:m_hat_j_expl_general} and \eqref{eq:m_bar_j_expl_general} with $k=0$.
The masses $\hat{m}$ and $\bar{m}$, and hence the corresponding probability distribution $\px^{\lambda}$, depend on $\lambda$ but this dependence is omitted when clear from context.
The transformation leading to the $N_r - 1$ difference equations \eqref{eq:result_diff_eq} removes the dependence on $I$ and $\cbar$.
In exchange, the cost constraint has to be imposed explicitly; earlier it was encoded implicitly in the set of equations \eqref{eq:eq_costr_IIb}.
As the mutual information $I$, the cost constraint depends on the entire input distribution.
However, by \lemref{chain_non_overlapping}, we know that there exists a unique Lagrange multiplier $\lambda^\ast$ such that the the cost constraint holds whenever our ansatz coincides with the capacity-achieving input distribution.

Knowing the values of the combined masses $\mhat{}$ and $\mbar{}$, we compute the original masses using the back transform \eqref{eq:def-masses-non-integer-case}.
A priori it is not guaranteed that $m_{j}>0$ for all $j$ independent of $\lambda$.
For $\lambda=0$, we obtain the masses (\ref{eq:def_masses_unconstr}) of the unconstrained Case I, where $m_{j}>0$ for all $j$.
The following lemma with $k=0$ guarantees that a solution with only positive weights exists also for increasing $\lambda>0$.
%%%%%%%%%%%%%%%%%%%%%%%%%% m_2 vanishes first %%%%%%%%%%%%%%%%%%%%%%%%%%
\begin{lemma}[$m_{2}$ vanishes first]
\label{lem:smallest-weight}
Let $c\pa{x}$ be a concave cost function as specified in \Cref{ass:cost_function} ($\alpha\leq1$) and $r\notin\mathbb{N}$.
Let the masses $m_j(\lambda)$, be defined by \eqref{eq:def-masses-non-integer-case}, where $\mhat{}$ and $\mbar{}$ are given by \eqref{eq:m_hat_j_expl_general} and \eqref{eq:m_bar_j_expl_general} with $k=0$.
Then there exists $\lambda_{0}>0$ such that
\begin{enumerate}[label=\arabic*)]
  \item $m_j(\lambda)>0$ for all $m_j\pa{\lambda}\in \Mkg{0}$ and all $\lambda\in[0,\lambda_{0})$;
  \item $m_{2}\pa{\lambda_{0}}=0$ and $m_j(\lambda_0)>0$ for all $m_j\in \Mkg{1}$, where, for $\alpha=1$, $\lambda_0\rightarrow\infty$.
\end{enumerate}
\end{lemma}
\begin{IEEEproof}
    We define $\lambda_0$ implicitly as the smallest $\lambda \geq 0$ for which
    \begin{equation}
        \zhat\pa{\lambda_0} = \zbar\pa{\lambda_0}\label{eq:def_lambda0}
    \end{equation}
    holds. Since
    \begin{equation}
    m_{2}       =  \mbar{1} - \mhat{1}      =  \zbar^{-1} - \zhat^{-1},
    \end{equation}
    this is equivalent to defining $\lambda_0$ as the smallest $\lambda$, where $m_{2}$ vanishes.
    For $\lambda = 0$ the normalization constants evaluate to
    $\zhat\pa{\lambda=0} = n+1$ and $\zbar\pa{\lambda=0}=n$, so that
    \begin{equation}
        \zhat\pa{\lambda}>\zbar\pa{\lambda},\;\lambda \in [0,\lambda_0).
    \end{equation}
    Note that by \lemref{properties_ds}, $\dhat{j}=\dbar{j}$ for $\alpha = 1$, so that the normalization constants satisfy $\zhat\pa{\lambda}=\zbar\pa{\lambda}+e^{-\lambda \dhat{n+1}}$, which implies $\lambda_0=\infty$.
    
    For $\alpha<1$, the difference $\zhat\pa{\lambda}-\zbar\pa{\lambda} \rightarrow -e^{\lambda \dbar{2}}<0$ for $\lambda \rightarrow \infty$ because it is dominated by the smallest exponent $\dbar{2}$. Hence, by the intermediate value theorem, there is a finite $\lambda_0$ that fulfills \eqref{eq:def_lambda0}.
    
    By \corref{smallest_masses}, the mass $m_{2}$ is the smallest of the even masses and $m_{2n-1}$ is the smallest of the odd masses.
    Using the definition \eqref{eq:def_mhat_mbar} of the combined masses, $m_{2n-1}$ is given by
    \begin{align}
        m_{2n-1} &= \mbar{n} - \mhat{n+1} \nonumber\\
        &= \zbar^{-1} e^{-\lambda \dbar{n}} - \zhat^{-1} e^{-\lambda \dhat{n+1}}.
    \end{align}
    If $\lambda \in [0,\lambda_0)$, it follows that $m_2>0$ because in this interval $\zhat > \zbar$.
    Moreover,
    \begin{equation}
    m_{2n-1}\geq \pa{\zbar^{-1} - \zhat^{-1}} e^{-\lambda \dhat{n+1}} >0,
    \end{equation}
    because $\dhat{n+1}>\dhat{n}>\dbar{n}$ by \lemref{properties_ds}.
    This proves statement \textit{1)}.\\
    If $\lambda=\lambda_0$, then $m_2=0$ but $m_{2n-1}$ remains positive because if we define $z\coloneqq \zhat\pa{\lambda_0}=\zbar\pa{\lambda_0}$, we obtain
    \begin{equation}
        m_{2n-1}=z^{-1}\pa{e^{-\lambda \dbar{n}}-e^{-\lambda \dhat{n+1}}}>0.
    \end{equation}
    Since all even masses are bigger than $m_2$ and all odd masses bigger than $m_{2n-1}$, this proves statement \textit{2)}.
\end{IEEEproof}
%%%%%%%%%%%%%%%%%%%%%%%%%%%%%%%%%%%%%%%%%%%%%%%%%%%%%%%%%%%%%%%%%%%%%%%%%
\begin{definition}
    \lemref{smallest-weight} defines $\lambda_{0}$ as the smallest value of $\lambda$ for which $x_2$ ceases to be a point of the support $S_0$ of our ansatz. 
    We define the corresponding costs as $\theta_0$, i.e.
    \begin{equation}
        \theta_0\coloneqq\Ex{c\pa{x}}_{\px^{\lambda_0}}.
    \end{equation}
\end{definition}
Hence, by definition, all masses $m_j$ are positive if $\cbar\in (\theta_0,\cbar^{\ast}]$. \figref{vanishing_masses} a) shows an example of how the masses behave as a function of $\cbar$.

As in Case IIa, we continue by showing that the Lagrange multiplier, which is the only remaining free variable of our ansatz, can be tuned to a value $\lambda^{\ast}$ such that the ansatz satisfies the equality constraint.
In contrast to the previous case, the following Lemma states only the existence of a $\lambda^{\ast}$.
%%%%%%%%%%%%%%%%%% ansatz fulfills equality constraint %%%%%%%%%%%%%%%%%%
\begin{lemma}
\label{lem:chain_overlapping_step_A}(Equality constraint, $r\notin\mathbb{N}$, Step A)
Let $c\pa{x}$ be a concave cost function as specified in \Cref{ass:cost_function} ($\alpha\leq1$), $\cbar \in (\theta_0,\bar c^{\ast}]$ and $r\notin\mathbb{N}$. Consider the discrete probability distribution $\px^{\lambda}$ parametrized as in \eqref{eq:general_ansatz}.
Let the positions $x_{j}$ be defined in \eqref{eq:def_pos_unconstr} and the masses $m_{j}\pa{\lambda}$ be defined by \eqref{eq:def-masses-non-integer-case} with $\mhat{}$ and $\mbar{}$ given by \eqref{eq:m_hat_j_expl_general} and \eqref{eq:m_bar_j_expl_general}, where $k=0$.
Then there is a $\lambda=\lambda^{\ast}$ so that $\px^{\lambda^{\ast}}$ is a solution to the equality constraint (\ref{eq:eq_constr}).
\end{lemma}
\begin{IEEEproof}
By construction, for a given $\lambda$, the $N_r$ masses $m_{j}$ satisfy the $N_r-1$ difference equations (\ref{eq:diff_eqs_integer_case}), which together with the cost constraint are equivalent to the $N_r$ original equations \eqref{eq:eq_costr_IIb} derived from the equality constraint \eqref{eq:eq_constr}.
Thus, it remains to be shown that there exists a $\lambda^{\ast}$ such that the cost constraint holds, i.e. $\Ex{c\pa{x}}_{\px^{\lambda^{\ast}}}=\cbar$.
Since the masses $m_j$ are continuous functions of $\lambda$, the cost $\Ex{c\pa{x}}_{\px^{\lambda}}$
Since our ansatz for $\px^{\lambda}$ with positions $x_j$ and masses $m_j$ as in the assumption of this lemma is a concatenation of continuous functions in $\lambda$, also the costs $\Ex{c\pa{x}}_{\px^{\lambda}}$ is a continuous function of $\lambda$, too.
For $\lambda=0$, we obtain the costs $\cbar^{\ast}$ of the unconstrained case and for $\lambda=\lambda_0$, we have $\cbar=\theta_0$ by definition.
Therefore, by the intermediate value theorem there exists a $\lambda^{\ast}$ such that $\Ex{c\pa{x}}_{\px^{\lambda}}=\cbar$.
\end{IEEEproof}
%%%%%%%%%%%%%%%%%%%%%%%%%%%%%%%%%%%%%%%%%%%%%%%%%%%%%%%%%%%%%%%%%%%%%%%%%
To conclude the proof of \emph{Step A}, we note that the same reasoning as in Case IIa applies.
The solution $\px^{\lambda^\ast}$ satisfies (\ref{eq:ineq_constr}) with equality at $x_{j}$, $j=1,\ldots,N_r$.
Moreover, \lemref{linear_i} shows that the l.h.s. of (\ref{eq:ineq_constr}) increases linearly between $x_{j}$ and $x_{j+1}$, and the r.h.s is concave due to $\alpha\leq1$.
Thus, (\ref{eq:ineq_constr}) is also satisfied for all the points $x\in\left(x_{j},x_{j+1}\right)$, see \figref{cost} a).
Hence, $\px^{\lambda^{\ast}}$ is the capacity-achieving input distribution $\px^{\ast}$ and its support is $S_0$, i.e. that of the unconstrained case.
By \lemref{chain_non_overlapping} the value of $\lambda^{\ast}$ is also unique.
%%%%%%%%%%%%%%%%%%%%%%%%%%%% summary step A %%%%%%%%%%%%%%%%%%%%%%%%%%%%%%

\emph{Summary of Step A}. With $r\notin \mathbb{N}$ fixed and a strictly convex cost function ($\alpha<1$), \lemref{smallest-weight} together with the expressions of the masses suggest the following qualitative behavior of the capacity-achieving input distribution for $\cbar \in (\theta_0,\cbar^\ast]$.
At $\cbar=\cbar^\ast$, the probability masses are given by \eqref{eq:def_masses_unconstr}.
A decreasing $\cbar^\ast$ redistributes the probability mass in a continuous manner between the same fixed positions. 
Since
\begin{align}
    \frac{\partial m_1 }{\partial \lambda} & =\frac{\partial}{\partial \lambda}z^{-1}=\frac{1}{u^2} \pa{\frac{\partial \mkl{k}}{\partial \lambda}\, u- \mkl{k}\,\frac{\partial u}{\partial \lambda}}\\
    & =\frac{1}{u^2} \pa{\frac{\partial \mkl{k}}{\partial \lambda}\, u + \mkl{k}\, \sum_{j=1}^{k}c_{2j-1}e^{-\lambda c_{2j-1}}}\\
    & >0,
\end{align}
where $\partial \mkl{k}/\partial \lambda >0$ by \lemref{properties_mkl_mkg}, more mass is transferred to the position with the lowest costs, i.e. to $x_1=0$.

The value of $m_2$ results from two conflicting effects.
On the one hand, $x_2$ has the second lowest costs but on the other hand, this input creates confusion with the most probable input at $x_1$.
As $\cbar \rightarrow \theta_{0}^{+}$, the negative impact of the latter effect on the mutual information prevails and the mass $m_2$ vanishes, while the others remain positive.

For the linear cost function ($\alpha=1$), $m_2$ does not vanish but all masses remain positive for all $\cbar\in(0, \cbar^\ast]$.

\subsubsection{Step B}
In \emph{Step B}, we describe the capacity-achieving input distribution for $\cbar\in (\theta_k,\theta_{k-1}]$.
The approach is similar to the previous proofs.
We guess an ansatz for $\px^{\ast}$ that by construction satisfies the equality constraint at the points of the support. 
Inspired by the extinction of $m_2$ as $\cbar\rightarrow \theta_{0}^{+}$, we drop the points $\pacu{x_{2j}\mid j=1,\,\ldots,\,k}$ from the support so that the corresponding masses $m_2,\ldots,m_{2k}$ are equal to zero and $S=S_k$ as defined in \eqref{eq:Sk_cases}.
The equality constraint at the remaining positions provides a set of equations that determines the distribution of the probability mass.
Similarly as before, we show that it is solved by non-negative weights if $\cbar\geq \theta_k$.
We then show that this ansatz for $\px^{\ast}$ also satisfies the inequality constraint if $\cbar\in (\theta_k,\theta_{k-1}]$.
Moreover, in the limit $\cbar\rightarrow\theta_{k}^{+}$ the solution converges to that of the limit $\cbar\rightarrow\theta_{k}^{-}$ of the next interval $\cbar\in (\theta_{k+1},\theta_{k}]$.
With this iterative argument, we cover $\cbar \in (\theta_{n-2},\theta_{0}]$.
%%%%%%%%%%%%%
% STEP B
%%%%%%%%%%%%

Consider our ansatz \eqref{eq:general_ansatz} for an input distribution $\px$ with support $S_k$ for some fixed $1\leq k \leq n-2$.
The resulting output distribution $\py$ can be considered as a combination of the integer and the non-integer case, see \figref{sketch_labeling}.
For the positions $x_j \in \xkl{k}$, the outputs $p_{N}\left(y - x_j\right)$ cover $y\in \pasq{-b,\,x_{2k-1}+b}$ without gap or overlap, similar to Case IIa, where $r\in \mathbb{N}$.
The other positions $x_j \in \xkg{k}$ lead to the overlapping outputs $p_{N}\left(y - x_j\right)$ for $y \in \pasq{x_{2k+1}-b,\,x_{2n}+b}$ considered in \emph{Step A}, where $r \notin \mathbb{N}$.
Therefore, the intuition is that the masses in $\Mkl{k}$ behave similarly to those of Case IIa and the remaining masses $m_j \in \Mkg{k}$ show the properties discussed in the previous step.

For fixed $k$, we start \emph{Step B} with a generalization of \lemref{smallest-weight}, which guarantees positive masses $m_j$ for $\lambda>\lambda_k$, where $\lambda_{k}>\lambda_{k-1}$.
%%%%%%%%%%%%%%%%%%%%%% m_2 vanishes first, Step B %%%%%%%%%%%%%%%%%%%%%%
\begin{lemma}[$m_{2k+2}$ vanishes first]
\label{lem:smallest-weight-step-b}
Let $c\pa{x}$ be a strictly convex cost function as specified in \Cref{ass:cost_function} ($\alpha<1$) and $r\notin\mathbb{N}$.
Let the masses $m_j(\lambda)$, be defined by \eqref{eq:def-masses-non-integer-case}, where $\mhat{}$ and $\mbar{}$ are given by \eqref{eq:m_hat_j_expl_general} and \eqref{eq:m_bar_j_expl_general} for some fixed $k\in\rangeone{n-2}$.
Then there exists $\lambda_{k}>\lambda_{k-1}$ such that
\begin{enumerate}[label=\arabic*)]
    \item $m_j(\lambda)>0$ for all $m_j\pa{\lambda}\in \Mkg{k}$ and all $\lambda\in[\lambda_{k-1},\lambda_{k})$;
    \item $m_{2k+2}\pa{\lambda_{k}}=0$ and $m_j(\lambda_k)>0$ for all $m_j\in \Mkg{k+1}$.
\end{enumerate}
Moreover, $m_j\pa{\lambda}>0$ for all $m_j\in\Mkl{k}$ and for all $\lambda>0$.
\end{lemma}
\begin{IEEEproof}
    \lemref{smallest-weight} proves the case $k=0$, which we use to build a recursive proof for all $k\in\rangeone{n-2}$.
    For a given $k>0$, we define $\lambda_k$ implicitly as the smallest $\lambda > \lambda_{k-1}$ for which
    \begin{equation}
        \zhat_{k}\pa{\lambda_k} = \zbar_{k}\pa{\lambda_k}\label{eq:def_lambdak}
    \end{equation}
    holds. In this proof, we denote explicitly the dependence of the definitions of $\zhat$ and $\zbar$ on $k$ in \eqref{eq:m_hat_j_expl_general} and \eqref{eq:m_bar_j_expl_general}.
    Since
    \begin{equation}
    m_{2k+2}       =  \mbar{k+1} - \mhat{k+1}      =  \zbar_{k}^{-1} - \zhat_{k}^{-1},
    \end{equation}
    this is equivalent to defining $\lambda_k$ as the smallest $\lambda>\lambda_{k-1}$, where $m_{2k+2}$ vanishes.
    The existence of such a $\lambda_{k}$ follows from evaluating the difference $\zhat_{k}\pa{\lambda}-\zbar_{k}\pa{\lambda}$ for $\lambda=\lambda_{k-1}$ and $\lambda\rightarrow\infty$.
    For $\lambda_{k-1}$ it holds that
    \begin{align}
    &&\zhat_{k-1}\pa{\lambda_{k-1}} & = \zbar_{k-1}\pa{\lambda_{k-1}}\\
    \Leftrightarrow&&\sum_{j=k+1}^{n+1}e^{-\lambda_{k-1} \dhat{j}^{k-1}} &= \sum_{j=k+1}^{n}e^{-\lambda_{k-1} \dbar{j}^{k-1}}\\
    \Rightarrow&&\sum_{j=k+2}^{n+1}e^{-\lambda_{k-1} \dhat{j}^{k-1}} &> \sum_{j=k+2}^{n}e^{-\lambda_{k-1} \dbar{j}^{k-1}}\\
    \Leftrightarrow&&\sum_{j=k+2}^{n+1}e^{-\lambda_{k-1} \pa{\dhat{j}^{k} + \dhat{k+1}^{k-1}}} &> \sum_{j=k+2}^{n}e^{-\lambda_{k-1} \pa{\dbar{j}^{k} + \dbar{k+1}^{k-1}}}\\
    \Leftrightarrow &&e^{-\lambda_{k-1}\dhat{k+1}^{k-1}}\,\zhat_{k}\pa{\lambda_{k-1}} & > e^{-\lambda_{k-1}\dbar{k+1}^{k-1}}\,\zbar_{k}\pa{\lambda_{k-1}}\\
    \Rightarrow &&\zhat_{k}\pa{\lambda_{k-1}} & > \zbar_{k}\pa{\lambda_{k-1}},
    \end{align}
    where the third and the last line follow from the fact that $\dhat{j}^{k}>\dbar{j}^{k}$ by \lemref{properties_ds}.
    
    For $\lambda \rightarrow \infty$, the difference $\zhat_{k}\pa{\lambda}-\zbar_{k}\pa{\lambda} \rightarrow -e^{-\lambda \dbar{k+2}}<0$ because the it is dominated by the exponent with the smallest absolute value, i.e. $-\lambda \dbar{k+2}$. Hence, by the intermediate value theorem there exists a finite $\lambda_k$ that fulfills \eqref{eq:def_lambdak}.
    
    By \lemref{properties_masses}, $m_{2n-1}$ is the smallest of the odd-indexed masses in $\Mkg{k}$.
    Using the definition \eqref{eq:def_mhat_mbar} of the combined masses, $m_{2n-1}$ is given by
    \begin{align}
        m_{2n-1} &= \mbar{n} - \mhat{n+1} \nonumber\\
        &= \zbar^{-1} e^{-\lambda \dbar{n}} - \zhat^{-1} e^{-\lambda \dhat{n+1}},
    \end{align}
    which is greater than zero because
    \begin{equation}
    m_{2n-1}\geq \pa{\zbar_{k}^{-1} - \zhat_{k}^{-1}} e^{-\lambda \dhat{n+1}} >0,
    \end{equation}
    where we used $\dhat{n+1}>\dhat{n}>\dbar{n}$ by \lemref{properties_ds}.
    Also by \lemref{properties_masses}, the mass $m_{2k+2}$ is the smallest of the even-indexed masses in $\Mkg{k}$ and for $\lambda \in [\lambda_{k-1},\lambda_{k})$, it follows that $m_{2k+2}>0$ because on this interval $\zhat_k > \zbar_k$, and
    \begin{equation}
        m_{2k+2}=\zbar_{k}^{-1}-\zhat_{k}^{-1}>0.
    \end{equation}
    This proves statement \textit{1)}.\\
    If $\lambda=\lambda_k$, then $m_{2k+2}=0$ but $m_{2n-1}$ remains positive because if we define $z_{k}\coloneqq \zhat_{k}\pa{\lambda_k}=\zbar_{k}\pa{\lambda_0}$, we obtain
    \begin{equation}
        m_{2n-1}=z_{k}^{-1}\pa{e^{-\lambda \dbar{n}}-e^{-\lambda \dhat{n+1}}}>0.
    \end{equation}
    Since all even-indexed masses in $\Mkg{k}$ are bigger than $m_{2k+2}$ and all odd masses in $\Mkg{k}$ are bigger than $m_{2n-1}$. This proves statement \textit{2)}.
\end{IEEEproof}
%%%%%%%%%%%%%%%%%%%%%%%%%%%%%%%%%%%%%%%%%%%%%%%%%%%%%%%%%%%%%%%%%%%%%%%%%
\begin{definition}
    \lemref{smallest-weight-step-b} defines $\lambda_{k}$ as the smallest $\lambda$, for which $x_{2k+2}$ ceases to be a point of the support $S_k$ of our ansatz.
    The corresponding costs define $\theta_{k}$, i.e.
    \begin{equation}
        \theta_{k}\coloneqq\Ex{c\pa{x}}_{\px^{\lambda_{k}}}.
    \end{equation}
\end{definition}
Hence, by definition, all masses $m_j\in \Mkg{k}$ are positive for $\cbar\in (\theta_{k},\theta_{k-1}]$.
This is similar to \emph{Step A}, where $m_j > 0$ is also guaranteed only on a finite interval for $\cbar$.
Similar to the integer case, the masses in $\Mkl{k}$ are positive for all $0<\cbar<\cbar^{\ast}$.

The following Lemma ensures that our ansatz satisfies the equality constraint on the interval $\cbar\in (\theta_{k},\theta_{k-1}]$.
\begin{definition}\label{def:theta_n}
    To include also the last interval $(0, \theta_{n-2}]$ while keeping the formulation of \lemref{chain_overlapping_step_B} and \lemref{ineq_at_x_2k} simple, we define
    \begin{equation}
        \theta_{n-1}\coloneqq 0,
    \end{equation}
    so that the last interval is equivalently denoted as $(\theta_{n-1}, \theta_{n-2}]$.
\end{definition}
%%%%%%%%%%%%%%% ansatz fulfills equality constraint, Step B %%%%%%%%%%%%%%
\begin{lemma}[Equality constraint, $r\notin\mathbb{N}$, Step B and C]
\label{lem:chain_overlapping_step_B}
Let $c\pa{x}$ be a strictly convex cost function as specified in \Cref{ass:cost_function} ($\alpha<1$), $r\notin\mathbb{N}$, and $\cbar \in (\theta_{k}, \theta_{k-1}]$, $k\in\rangeone{n}$. Let $\px^{\lambda}$ be the discrete probability distribution \eqref{eq:general_ansatz} with the positions $x_{j}$ as defined in \eqref{eq:def_pos_unconstr} and the masses $m_{j}\pa{\lambda}$ as defined in \eqref{eq:def-masses-non-integer-case}, with $\mhat{}$ and $\mbar{}$ given by \eqref{eq:m_hat_j_expl_general} and \eqref{eq:m_bar_j_expl_general}.
Then there is a $\lambda=\lambda^{\ast}$ so that $\px^{\lambda^{\ast}}$ is a solution to the equality constraint (\ref{eq:eq_constr}).
\end{lemma}
\begin{IEEEproof}
Consider the parametrization \eqref{eq:general_ansatz}, where the positions are given by $S_{k}$ for some fixed $k$ and the masses as described in \thmref{main} Case IIb.
We denote it as $p_{x,k}^{\lambda}$ in this section to make the dependence on $k$ explicit.

At the positions $x_j \in \xkl{k}$, the marginal information evaluates to $i\pa{x_{j};p_{x,k}^{\lambda}}=-\log m_{j}$, so that the equality constraint (\ref{eq:eq_constr}) can be written as
\begin{equation}
-\log m_{2j-1}=I+\lambda\left(c_{2j-1}-\bar{c}\right),\;j\in\rangeone{k}.\label{eq:eq_constr_x_smaller}
\end{equation}
Similarly to the proof of Case IIa, see Eq. \eqref{eq:diff_eqs_integer_case}, taking the difference between \eqref{eq:eq_constr_x_smaller} evaluated at $j$ and $j+1$ yields $k-1$ equations of the form
\begin{equation}
m_{2j+1}=m_{2j-1}\,e^{-\lambda\left(c_{2j+1}-c_{2j-1}\right)}.
\end{equation}
Requesting the normalization $\sum_{m_j\in \Mkl{k}} \,m_j = \mkl{k}$ yields the expressions in \eqref{eq:m_smaller}.

At the positions $x_j \in \xkg{k}$, the marginal information density evaluates to
\begin{equation}
i\left(x_{j},p_{x,k}^{\lambda}\right) =-\rho\log\hat{m}_{\hat{f}\left(j\right)}-\left(1-\rho\right)\log\bar{m}_{\bar{f}\left(j\right)},\label{eq:i_greater}
\end{equation}
where the labels are given by $\hat{f}\pa{j}$ and $\bar{f}\pa{j}$ as defined in \eqref{eq:def_fhat_fbar}.
Inserting (\ref{eq:i_greater}) into the equality constraint (\ref{eq:eq_constr}), subtracting the $\left(2j\right)$-th equality from the $\left(2j-1\right)$-th equality, and subtracting the $\left(2j+1\right)$-th equality from the $\left(2j\right)$-th equality gives $2\pa{n-k}+1$ difference equations that can be expresses as
\begin{align}
\hat{m}_{j+1} & =\hat{m}_{j}e^{-\lambda\frac{\widehat{\Delta c}_{j}}{\rho}},\;j\in\range{k+1}{n+1}\\
\bar{m}_{j+1} & =\bar{m}_{j}e^{-\lambda\frac{\overline{\Delta c}_{j}}{1-\rho}},\;j\in\range{k+1}{n},
\end{align}
respectively.
The factors $\widehat{\Delta c}_{j}$ and $\overline{\Delta c}_{j}$ are defined as in \eqref{eq:def_chat_cbar}.
Using the definitions of $\dhat{j}$ and $\dbar{j}$, see  Eqs.~\eqref{eq:definition_dhat_general} and \eqref{eq:definition_dbar_general}, and requesting the normalization 
\begin{equation}
    \sum_{j=k+1}^{n+1}\mhat{j}=\sum_{j=k+1}^{n}\mbar{j}=\mkg{k},\label{eq:norm_hat_bar}
\end{equation}
we obtain \eqref{eq:m_hat_j_expl_general} and \eqref{eq:m_bar_j_expl_general}.

Taking the difference between \eqref{eq:eq_constr} evaluated at $x_{1}$ and at $x_{2k+1}$, i.e. the leftmost positions in $\xkl{k}$ and $\xkg{k}$, respectively, yields one additional difference equation, that determines the ratio $\mkl{k}\pa{\lambda}/\mkg{k}\pa{\lambda}$.
Hence, in total, we have $2n-k-1$ difference equations, which together with the cost constraint are equivalent to the $2n-k$ equations obtained from the equality constraint \eqref{eq:eq_constr}.
By construction, our ansatz $\px^{\lambda}$ satisfies the difference equations for all $\lambda$, so that it also satisfies the equality constraint if
\begin{enumerate}
    \item there is a $\lambda^{\ast}$ such that the cost constraint holds;
    \item our ansatz yields positive values for the masses for $\cbar \in (\theta_{k}, \theta_{k-1}]$.
\end{enumerate}
The existence of $\lambda^{\ast}$ such that the costs do not exceed $\theta_{k-1}$ is guaranteed because for $\lambda\rightarrow\infty$, the ratio $\mkl{k}/\mkg{k}\rightarrow\infty$, by \lemref{properties_mkl_mkg}, so that all the probability mass is located in $\Mkl{k}$.
Within $\Mkl{k}$, all the probability mass is concentrated at $m_1$, so that we have vanishing costs in the limit.

Positivity of the weights follow from the following observation.
Due to the exponential form, the masses $m_j \in \Mkl{k}$ are always positive, so that only the masses $m_j\in \Mkg{k}$ can potentially be negative.
Define $\lambda_{k-1}$ as the value of the Lagrange multiplier where the mass $m_{2k}$ of the ansatz $p_{x,k-1}^{\lambda}$ with $x_j\in \xkg{k-1}$ and $m_j\in\Mkg{k-1}$ vanishes.
At $\lambda_{k-1}$, the two ans\"{a}tze $p_{x, k}^{\lambda_{k-1}}$ and $p_{x, k-1}^{\lambda_{k-1}}$ are equal because they satisfy the same difference equations and $m_{2k}=0$.
Since $m_{2k}$ vanishes at $\lambda_{k-1}$, we know by \lemref{smallest-weight-step-b} applied to $p_{x,k-1}^{\lambda}$, that all masses $m_j\in\Mkg{k}$ are positive.
By \lemref{smallest-weight-step-b} applied to $p_{x,k}^{\lambda}$, the mass $m_{2k+2}$ vanishes next and we define the corresponding costs as $\theta_{k}$.

Hence, the ansatz \eqref{eq:general_ansatz} with support $S_{k}$ satisfies the equality constraint and has positive masses for $\cbar \in (\theta_{k},\theta_{k-1}]$.
\end{IEEEproof}
%%%%%%%%%%%%%%%%%%%%%%%%%%%%%%%%%%%%%%%%%%%%%%%%%%%%%%%%%%%%%%%%%%%%%%%%%
It remains to be shown that our ansatz with $\lambda=\lambda^{\ast}$ also meets the inequality constraint.
Intuitively, the constraint has to hold on $x\in \pasq{x_0,x_{2k-1}}$ for the same reasons as it is satisfied in Case IIa, and the for $x \in \pasq{x_{2k+1},x_{2n}}$, the same reasoning as in \emph{Step A} applies.
Only the intermediate interval $x \in \pa{x_{2k-1}, x_{2k+1}}$ requires a more careful investigation, which is provided by the following Lemma.
%%%%%%%%%%%%%% ansatz fulfills inequality constraint, Step B %%%%%%%%%%%%%
\begin{lemma}[Inequality constraint for $r\notin\mathbb{N}$, Step B and C]
\label{lem:ineq_at_x_2k}
Let $c\pa{x}$ be a strictly convex cost function as specified in \Cref{ass:cost_function} ($\alpha<1$), $\bar{c}\in(\theta_{k},\theta_{k-1}]$ for some fixed $k\in\rangeone{n}$ and let $r\notin\mathbb{N}$. Further, let $\px^{\lambda}$ be the discrete probability distribution \eqref{eq:general_ansatz}.
Let the positions $x_{j}$ be defined in \eqref{eq:def_pos_unconstr} and the masses $m_{j}\pa{\lambda}$ be defined by \eqref{eq:def-masses-non-integer-case} with $\mhat{}$ and $\mbar{}$ given by \eqref{eq:m_hat_j_expl_general} and \eqref{eq:m_bar_j_expl_general}.
Then there is a $\lambda=\lambda^{\ast}$ so that $\px^{\lambda^{\ast}}$ is a solution to the inequality constraint (\ref{eq:ineq_constr}).
\end{lemma}
\begin{IEEEproof}
The positions $x_j\in\xkl{k}$ are given by \eqref{eq:def_pos_unconstr} for $r\notin \mathbb{N}$ and $j$ odd. Therefore, \lemref{linear_i} applies with the relabeling $x_{2k+1}\rightarrow x_k$, i.e. as in the case $r\in \mathbb{N}$. Thus, the marginal information density $i\pa{x;\px^{\ast}}$ interpolates linearly between the $x_j$ for $x\in\pasq{x_1=0,\,x_{2k-1}}$ and due to the concavity of the cost function, the inequality constraint \eqref{eq:ineq_constr} is fulfilled on this interval.
Similarly, the positions $x_j\in \xkg{k}$ are those of \eqref{eq:def_pos_unconstr} for $r\in \mathbb{N}$. Hence, \lemref{linear_i} also applies to these positions which proves \eqref{eq:ineq_constr} for $x\in \pasq{x_{2k+1},\,x_{2n}=1}$.

To prove that \eqref{eq:ineq_constr} is satisfied on the remaining interval $x\in\pasq{x_{2k-1},\,x_{2k+1}}$, we first show in \ref{itm:kink} that $i\pa{x;\px^{\ast}}$ is linear on each of the two subintervals $\pasq{x_{2k-1},\,x_{2k}}$ and $\pasq{x_{2k},\,x_{2k+1}}$. In \ref{itm:ineq_x_2k} we then show that \eqref{eq:ineq_constr} is fulfilled at $x=x_{2k}$ and therefore on the entire interval $x\in\pasq{x_{2k-1},\,x_{2k+1}}$.
%%%%%%%%%%%%%%%%%%%%%%%%%%%%%%%%%%%%%%%%%%%%%%%%%%%%%%%%%%%%%%%%%%%%%%%%%
\begin{figure}
\centerline{\includegraphics[width=0.75\figW]{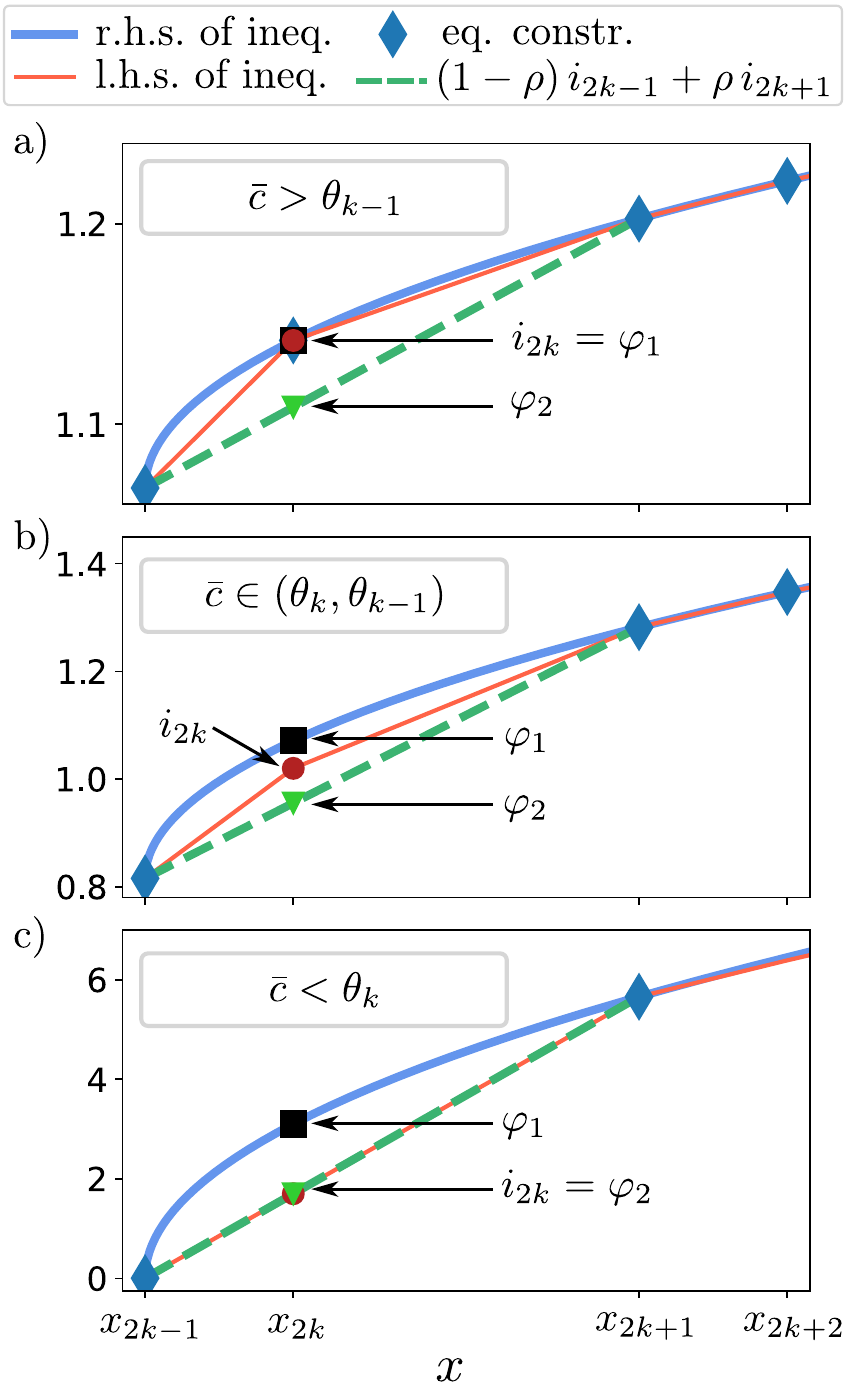}}
\caption{Illustration of the satisfaction of the inequality constraint at $x_{2k}$, with $k=1$ chosen as an example. Solid curves denote the l.h.s. (blue) and the r.h.s. (red) of the inequality constraint \eqref{eq:ineq_constr} as a function of $x$ for different values of $\cbar$. The blue diamonds indicate the positions where the equality constraint \eqref{eq:eq_constr} holds. The y-values of the black square, the red dot, and the green triangle indicate the values of $\phi_{1}$, $i_{2k}$, and $\phi_{2}$, respectively. Here, $r=2.3$, so that there are positions $x_1,\,\ldots,\,x_6$. The two last positions $x_5$ and $x_6$ are outside of the range of the x-axis. a) $\cbar>\theta_{k-1}$, where $m_{2k}>0$ and therefore $x_{2k}\in S$, so that the equality constraint holds, i.e. $i_{2k}=\phi_{1}$. b) $\cbar\in(\theta_{k},\theta_{k-1})$, where $m_{2k}=0$ but $m_{2k+2}>0$. c) $\cbar<\theta_{k}$, where $m_{2k}=m_{2k+2}=0$. Therefore, $\xkl{k}=\pacu{x_1,\,x_3,\,\ldots,\,x_{2k-1}}$, so that $i_{2k}$ coincides with $\phi_2$. \vspace{-0.3cm}
\label{fig:ineq_x_2k}}
\end{figure}
%%%%%%%%%%%%%%%%%%%%%%%%%%%%%%%%%%%%%%%%%%%%%%%%%%%%%%%%%%%%%%%%%%%%%%%%%
\begin{enumerate}[label=\textit{\arabic*)},leftmargin=*,itemsep=0.25em,topsep=0.25em]
\item \label{itm:kink}The marginal information density on $\pasq{x_{2k-1},\,x_{2k}}$ evaluates to
\begin{align}
    i\left(x;\px\right) & =-\frac{1}{2b}\int_{x-b}^{x+b}dy\,\log\left[2b\,\py^{\ast}\left(y\right)\right]\\
    &=-\frac{1}{2b}\Bigg[\int_{x-b}^{x_{2k-1}+b}dy\,\log m_{2k-1}\nonumber\\
    &\hphantom{=-\frac{1}{2b}\Bigg[}\;{}+\int_{x_{2k-1}+b}^{x+b}dy\,\log m_{2k+1}\Bigg]\\
    &=d\,\log\frac{m_{2k-1}}{m_{2k+1}} - \log m_{2k-1},
\end{align}
where we introduced $d\coloneqq\pa{x-x_{2k-1}}/\pa{2b}$ to arrive at the last line, which shows the linearity of $i\pa{x;\px^{\ast}}$ in $x$ on the first interval.
For $x\in \pasq{x_{2k},\,x_{2k+1}}$, we obtain
\begin{align}
    i\left(x;\px\right) & = -\frac{1}{2b}\Bigg[\int_{x-b}^{x_{2k-1}+b}dy\,\log m_{2k-1}\nonumber\\
    &\hphantom{=-\frac{1}{2b}\Bigg[}\;{}+\int_{x_{2k-1}+b}^{x_{2k}+b}dy\,\log m_{2k+1}\nonumber\\
    &\hphantom{=-\frac{1}{2b}\Bigg[}\;{}+\int_{x_{2k}+b}^{x+b}dy\,\log\pa{m_{2k+1}+m_{2k+2}}\Bigg]\\
    & = d\,\log\frac{m_{2k-1}}{m_{2k+1}+m_{2k+2}} + \rho \log\frac{m_{2k-1}}{m_{2k+1}}\nonumber\\
    &\hphantom{=}\;{}- \log \frac{m_{2k-1}}{m_{2k+1}+m_{2k+2}},
\end{align}
with the definitions $d\coloneqq \pa{x-x_{2k}}/\pa{2b}$, and $\rho = r-\floor{r}$.
Thus, $i\pa{x,\px^{\ast}}$ is also linear on the second interval.
\item \label{itm:ineq_x_2k}For a given $k$, the inequality constraint is fulfilled with equality at $x_{2k-1}$ and $x_{2k+1}$ since both points belong to the support $S_k$, regardless of the value of $k$.
We know that $x_{2k}$ belongs to $S_{k-1}$ because for $\cbar>\theta_{k-1}$, the mass $m_{2k}>0$.
Hence, the equality constraint \eqref{eq:eq_constr} is satisfied, see \figref{ineq_x_2k} a).
Defining
\begin{equation}
    \varphi_1\pa{\cbar}\coloneqq I+\lambda\pa{c_{2k}-\cbar},
\end{equation}
we therefore know
\begin{equation}
    i_{2k}\pa{\cbar}=\varphi_1\pa{\cbar_1}\quad \mathrm{for}\;\cbar>\theta_{k-1},\label{eq:phi1}
\end{equation}
where we introduced the shorthand notation $i_{2k}\pa{\cbar_1}\coloneqq i\pa{x_{2k};\px^{\lambda^{\ast}\pa{\cbar}}}$.
As soon as $\cbar\leq\theta_k$, where $m_{2k+2}$ vanishes, the kink of $i_{2k}$ vanishes, too, and the marginal information density is given by a line for $x\in \pasq{x_{2k-1},x_{2k+1}}$, see \figref{ineq_x_2k} c). This is expected because both, $x_{2k-1}$ and $x_{2k+1}$ belong to $\xkl{k+1}$, where the reasoning of step \ref{itm:kink} applies. Defining
\begin{align}
    \varphi_2\pa{\cbar}&\coloneqq \pa{1-\rho} \,i_{2k-1}\pa{\cbar} + \rho\,i_{2k+1}\pa{\cbar}\\
    &=\pa{1-\rho} \,\pasq{I+\lambda \pa{c_{2k-1}-\cbar}}\nonumber\\
    &\hphantom{=}\;{}+ \rho\,\pasq{I+\lambda \pa{c_{2k+1}-\cbar}},
\end{align}
we can write
\begin{equation}
    i_{2k}\pa{\cbar}=\varphi_2\pa{\cbar}\quad \mathrm{for}\;\cbar_2\leq\theta_{k}.\label{eq:phi2}
\end{equation}
Since the inequality is fulfilled for \eqref{eq:phi1} and \eqref{eq:phi2}, our aim is to show that $i_{2k}\pa{\cbar}$ moves from $\varphi_1$ to $\varphi_2$ without exceeding $\varphi_1$ as $\cbar$ decreases from $\theta_{k-1}$ to $\theta_k$, see \figref{ineq_x_2k} b).
In the following calculations, we will use that within our parametrization of $\px$,
%, by \lemref{chain_non_overlapping}, 
$\cbar\pa{\lambda}=\Ex{c\pa{x}}_{\px^{\lambda}}$ is a continuous function of $\lambda$
%and strictly decreasing function 
and consider all quantities as functions of $\lambda$. The difference $\Delta\pa{\rho,\lambda}\coloneqq \varphi_1-\varphi_2$ can be written as
\begin{align}
    \Delta\pa{\rho,\lambda}&=\lambda\Big[c_{2k}-\pa{\rho\, c_{2k+1}+\pa{1-\rho}\,c_{2k-1}}\Big]\\
    &\geq 0,\quad \mathrm{for}\;\lambda\geq 0,
\end{align}
where we used the concavity of the cost function in the second line. This relation shows that $\Delta\pa{\rho,\lambda}$ increases linearly with lambda. Hence, to show that $i_{2k}$ will not exceed $\varphi_1$, it is sufficient to show that
\begin{equation}
    \frac{\partial}{\partial\lambda}\pa{i_{2k}-\varphi_2}<0,\label{eq:ineq_x_2k_to_show}
\end{equation}
because if $i_{2k}$ moves closer to $\varphi_2$ with increasing $\lambda$, it will never reach $\varphi_1$, which moves away from $\varphi_2$. Using
\begin{align}
    i_{2k-1}& = -\log m_{2k-1},\\
    i_{2k} & = -\rho\log m_{2k+1} - \pa{1-\rho}\log m_{2k-1},\\
    i_{2k+1} & = -\rho\log m_{2k+1} \nonumber\\
    &\hphantom{=\;{}}- \pa{1-\rho}\log\pa{m_{2k+1}+m_{2k+2}},
\end{align}
for $\cbar\in [\theta_k,\theta_{k-1})$, yields
\begin{equation}
    i_{2k}-\varphi_2 = \rho \pa{1-\rho}\log\pa{1+\frac{m_{2k+2}}{m_{2k+1}}}.
\end{equation}
Since only the masses depend on $\lambda$ and $\rho\pa{1-\rho}>0$, we will show
\begin{equation}
    \frac{\partial}{\partial \lambda}\frac{m_{2k+2}}{m_{2k+1}}<0.
\end{equation}
Inserting the expressions for the masses \eqref{eq:m_hat_j_expl_general}-\eqref{eq:m_bar_j_expl_general}, we obtain $m_{2k+2}/m_{2k+1}=\zhat/\zbar-1$, and therefore
\begin{align}
    \frac{\partial}{\partial \lambda}\frac{m_{2k+2}}{m_{2k+1}} &= \frac{\partial}{\partial \lambda} \frac{\zhat}{\zbar}=\frac{\partial}{\partial \lambda} \frac{\zhat^{\prime}}{\zbar^{\prime}}\\
    &=\frac{\zhat^{\prime}}{\zbar^{\prime}}\pasq{\zhat^{\prime-1}\frac{\partial\zhat^{\prime}}{\partial\lambda}-\zbar^{\prime-1}\frac{\partial\zbar^{\prime}}{\partial\lambda}}\\
    &=\frac{\zhat^{\prime}}{\zbar^{\prime}}\pasq{\frac{\partial}{\partial\lambda}\log\zhat^{\prime}-\frac{\partial}{\partial\lambda}\log\zbar^{\prime}}\\
    &=\frac{\zhat^{\prime}}{\zbar^{\prime}}\pasq{\Ex{\dbar{}}_{\mbar{}^{\prime}}-\Ex{\dhat{}}_{\mhat{}^{\prime}}},
\end{align}
where we used the rescaled normalizations $\zhat^{\prime}$ and $\zbar^{\prime}$ and the expectation values with respect to the Boltzmann distributions introduced in \Cref{rem:boltzmann_interpretation}. Since $\zhat^{\prime}/\zbar^{\prime}>0$, we continue with
\begin{align}
    &\Ex{\dbar{}}_{\mbar{}^{\prime}}-\Ex{\dhat{}}_{\mhat{}^{\prime}}\\
    &=  \sum_{j=k+1}^{n}\mbar{j}^{\prime}\dbar{j} - \sum_{j=k+1}^{n+1}\mhat{j}^{\prime}\dhat{j}\\
    &=\sum_{j=k+1}^{n}\pasq{\mbar{j}^{\prime}\dbar{j} - \mhat{j}^{\prime}\dhat{j}} - \mhat{n+1}^{\prime}\,\dhat{n+1}\\
    &=\sum_{j=k+1}^{n}\pasq{\mbar{j}^{\prime}\dbar{j} - \mhat{j}^{\prime}\dhat{j}-\dhat{n+1}\pa{\mbar{j}^{\prime}-\mhat{j}^{\prime}}}\\
    &<\sum_{j=k+1}^{n}\pasq{\dhat{j}\pa{\mbar{j}^{\prime}-\mhat{j}^{\prime}}-\dhat{n+1}\pa{\mbar{j}^{\prime}-\mhat{j}^{\prime}}}\\
    &<\dhat{n+1}\sum_{j=k+1}^{n}\pasq{\pa{\mbar{j}^{\prime}-\mhat{j}^{\prime}}-\pa{\mbar{j}^{\prime}-\mhat{j}^{\prime}}}=0.
\end{align}
In the third line, we used that the masses $\mhat{j}^{\prime}$ sum to one, so that
\begin{align}
    \mhat{n+1}&=1-\sum_{j=k+1}^{n}\mhat{j}\\
    &= \sum_{j=k+1}^{n}\pa{\mbar{j}-\mhat{j}}.
\end{align}
Moreover, by \lemref{properties_ds}, we used $\dhat{j}>\dbar{j}$ in the fourth and $\dhat{n+1}>\dhat{j}$, $j<n+1$, in the last line.
Therefore, we conclude that \eqref{eq:ineq_x_2k_to_show} is fulfilled and $i\pa{x_{2k};\px^\ast}\leq I+\lambda^{\ast}\pa{c_{2k}-\cbar}$.
\end{enumerate}
\end{IEEEproof}
Since the ansatz with support $S_k$ satisfies the two necessary and sufficient conditions and all masses are positive for $\bar{c}\in(\theta_{k},\theta_{k-1}]$, we conclude that this is the capacity-achieving input distribution on this interval.
Therefore, an appropriate $\lambda^{\ast}$ exists for every required $\cbar$ and by \lemref{chain_non_overlapping}, its value is unique.
Tightening the cost budget $\cbar$, we can apply the reasoning of \emph{Step B} repeatedly, dropping $m_{2k+2}$ from the support whenever we reach $\cbar=\theta_{k}$, until $\cbar=\theta_{n-2}$. 

\subsubsection{Step C}
In the last Step, we consider the final interval $\cbar \in (0,\theta_{n-1}]$ separately.
\lemref{chain_overlapping_step_B} and \lemref{ineq_at_x_2k}, that prove the satisfaction of the equality and the inequality constraints \eqref{eq:eq_constr} and \eqref{eq:ineq_constr}, remain valid also on this interval, see \Cref{def:theta_n}.
Only \lemref{smallest-weight-step-b} does not hold for $k=n-1$ and needs to be adapted as follows.
\begin{lemma}[$m_{2n}$ does not vanish]
    Let $c\pa{x}$ be a strictly convex cost function as specified in \Cref{ass:cost_function} ($\alpha<1$).
    Let the masses $m_j(\lambda)$, be defined by \eqref{eq:def-masses-non-integer-case}, where $\mhat{}$ and $\mbar{}$ are given by \eqref{eq:m_hat_j_expl_general} and \eqref{eq:m_bar_j_expl_general} with $k=n-1$.
    Then $m_{j}>0$ for all $m_j$ in $\Mkl{n-1}$ and in $\Mkg{n-1}$.
\end{lemma}
\begin{IEEEproof}
    For $k=n-1$, we obtain $\Mkg{n-1}=\pacu{m_{2n-1},\,m_{2n}}$, so that
\begin{align}
    m_{2n-1} &= \mhat{n} = \zhat^{-1},\\
    m_{2n} &= \mhat{n+1} = \zhat^{-1}e^{-\lambda \frac{c_{2n}-c_{2n-1}}{\rho}},\\
    \zhat &= 1+e^{-\lambda\frac{c_{2n}-c_{2n-1}}{\rho}}
\end{align}
where $\rho\coloneqq r-n+1$ as before.
Clearly, these expressions are positive regardless of the value of $\lambda$.
The same is true for the exponential expressions for the masses $m_j\in \Mkl{n-1}$, as defined in \eqref{eq:m_smaller}.
\end{IEEEproof}
This concludes \emph{Step C} and, hence, the proof of Case IIb.

\section*{Acknowledgment}

The authors would like to thank Hui-An Shen for fruitful discussions.
All authors received funding from the SNF grant 310030\_212247 “Why Spikes?”

\ifCLASSOPTIONcaptionsoff
  \newpage
\fi

\bibliographystyle{IEEEtran}
\bibliography{uniform_noise_extended}

\begin{thebibliography}{10}
\providecommand{\url}[1]{#1}
\csname url@samestyle\endcsname
\providecommand{\newblock}{\relax}
\providecommand{\bibinfo}[2]{#2}
\providecommand{\BIBentrySTDinterwordspacing}{\spaceskip=0pt\relax}
\providecommand{\BIBentryALTinterwordstretchfactor}{4}
\providecommand{\BIBentryALTinterwordspacing}{\spaceskip=\fontdimen2\font plus
\BIBentryALTinterwordstretchfactor\fontdimen3\font minus
  \fontdimen4\font\relax}
\providecommand{\BIBforeignlanguage}[2]{{%
\expandafter\ifx\csname l@#1\endcsname\relax
\typeout{** WARNING: IEEEtran.bst: No hyphenation pattern has been}%
\typeout{** loaded for the language `#1'. Using the pattern for}%
\typeout{** the default language instead.}%
\else
\language=\csname l@#1\endcsname
\fi
#2}}
\providecommand{\BIBdecl}{\relax}
\BIBdecl

\bibitem{shannon_mathematical_1948}
C.~E. Shannon, ``\BIBforeignlanguage{en}{A {Mathematical} {Theory} of
  {Communication}},'' \emph{\BIBforeignlanguage{en}{The Bell system technical
  journal}}, vol.~27, no.~3, pp. 379--423, 1948.

\bibitem{smith_information_1971}
J.~G. Smith, ``\BIBforeignlanguage{en}{The {Information} {Capacity} of
  {Amplitude}- and {Variance}-{Constrained} {Scalar} {Gaussian} {Channels}},''
  \emph{\BIBforeignlanguage{en}{Information and Control}}, vol.~18, pp.
  203--219, 1971.

\bibitem{oettli_capacity-achieving_1974}
\BIBentryALTinterwordspacing
W.~Oettli, ``\BIBforeignlanguage{en}{Capacity-achieving input distributions for
  some amplitude-limited channels with additive noise ({Corresp}.)},''
  \emph{\BIBforeignlanguage{en}{IEEE Transactions on Information Theory}},
  vol.~20, no.~3, pp. 372--374, May 1974. [Online]. Available:
  \url{http://ieeexplore.ieee.org/document/1055225/}
\BIBentrySTDinterwordspacing

\bibitem{shamai_capacity_1995}
\BIBentryALTinterwordspacing
S.~Shamai and I.~Bar-David, ``\BIBforeignlanguage{en}{The capacity of average
  and peak-power-limited quadrature {Gaussian} channels},''
  \emph{\BIBforeignlanguage{en}{IEEE Transactions on Information Theory}},
  vol.~41, no.~4, pp. 1060--1071, Jul. 1995. [Online]. Available:
  \url{http://ieeexplore.ieee.org/document/391243/}
\BIBentrySTDinterwordspacing

\bibitem{lapidoth_capacity_2009}
\BIBentryALTinterwordspacing
A.~Lapidoth and S.~M. Moser, ``On the {Capacity} of the {Discrete}-{Time}
  {Poisson} {Channel},'' \emph{IEEE Transactions on Information Theory},
  vol.~55, no.~1, pp. 303--322, Jan. 2009. [Online]. Available:
  \url{http://ieeexplore.ieee.org/document/4729780/}
\BIBentrySTDinterwordspacing

\bibitem{dytso_when_2018}
\BIBentryALTinterwordspacing
A.~Dytso, M.~Goldenbaum, H.~V. Poor, and S.~S. Shitz,
  ``\BIBforeignlanguage{en}{When are discrete channel inputs optimal? —
  {Optimization} techniques and some new results},'' in
  \emph{\BIBforeignlanguage{en}{2018 52nd {Annual} {Conference} on
  {Information} {Sciences} and {Systems} ({CISS})}}.\hskip 1em plus 0.5em minus
  0.4em\relax Princeton, NJ: IEEE, Mar. 2018, pp. 1--6. [Online]. Available:
  \url{https://ieeexplore.ieee.org/document/8362306/}
\BIBentrySTDinterwordspacing

\bibitem{dytso_capacity_2019}
\BIBentryALTinterwordspacing
A.~Dytso, M.~Al, H.~V. Poor, and S.~Shamai~Shitz, ``\BIBforeignlanguage{en}{On
  the {Capacity} of the {Peak} {Power} {Constrained} {Vector} {Gaussian}
  {Channel}: {An} {Estimation} {Theoretic} {Perspective}},''
  \emph{\BIBforeignlanguage{en}{IEEE Transactions on Information Theory}},
  vol.~65, no.~6, pp. 3907--3921, Jun. 2019. [Online]. Available:
  \url{https://ieeexplore.ieee.org/document/8598797/}
\BIBentrySTDinterwordspacing

\bibitem{dytso_capacity_2020}
\BIBentryALTinterwordspacing
A.~Dytso, S.~Yagli, H.~V. Poor, and S.~Shamai~Shitz,
  ``\BIBforeignlanguage{en}{The {Capacity} {Achieving} {Distribution} for the
  {Amplitude} {Constrained} {Additive} {Gaussian} {Channel}: {An} {Upper}
  {Bound} on the {Number} of {Mass} {Points}},''
  \emph{\BIBforeignlanguage{en}{IEEE Transactions on Information Theory}},
  vol.~66, no.~4, pp. 2006--2022, Apr. 2020. [Online]. Available:
  \url{https://ieeexplore.ieee.org/document/8878162/}
\BIBentrySTDinterwordspacing

\bibitem{eisen_capacity-achieving_2023}
\BIBentryALTinterwordspacing
J.~Eisen, R.~R. Mazumdar, and P.~Mitran,
  ``\BIBforeignlanguage{en}{Capacity-{Achieving} {Input} {Distributions} of
  {Additive} {Vector} {Gaussian} {Noise} {Channels}: {Even}-{Moment}
  {Constraints} and {Unbounded} or {Compact} {Support}},''
  \emph{\BIBforeignlanguage{en}{Entropy}}, vol.~25, no.~8, p. 1180, Aug. 2023.
  [Online]. Available: \url{https://www.mdpi.com/1099-4300/25/8/1180}
\BIBentrySTDinterwordspacing

\bibitem{barletta_binomial_2024}
\BIBentryALTinterwordspacing
L.~Barletta, I.~Zieder, A.~Favano, and A.~Dytso,
  ``\BIBforeignlanguage{en}{Binomial {Channel}: {On} the {Capacity}-{Achieving}
  {Distribution} and {Bounds} on the {Capacity}},'' in
  \emph{\BIBforeignlanguage{en}{2024 {IEEE} {International} {Symposium} on
  {Information} {Theory} ({ISIT})}}.\hskip 1em plus 0.5em minus 0.4em\relax
  Athens, Greece: IEEE, Jul. 2024, pp. 711--716. [Online]. Available:
  \url{https://ieeexplore.ieee.org/document/10619601/}
\BIBentrySTDinterwordspacing

\bibitem{fahs_2018}
J.~Fahs and I.~Abou-Faycal, ``On properties of the support of
  capacity-achieving distributions for additive noise channel models with input
  cost constraints,'' \emph{IEEE Transactions on Information Theory}, vol.~64,
  no.~2, pp. 1178--1198, 2018.

\bibitem{dytso_poisson_2021}
A.~Dytso, L.~Barletta, and S.~Shamai~Shitz, ``Properties of the support of the
  capacity-achieving distribution of the amplitude-constrained {Poisson} noise
  channel,'' \emph{IEEE Transactions on Information Theory}, vol.~67, no.~11,
  pp. 7050--7066, 2021.

\bibitem{fahs_2014}
J.~Fahs and I.~Abou-Faycal, ``A {Cauchy} input achieves the capacity of a
  {Cauchy} channel under a logarithmic constraint,'' in \emph{2014 IEEE
  International Symposium on Information Theory}, 2014, pp. 3077--3081.

\bibitem{anantharam_1996}
V.~Anantharam and S.~Verdu, ``Bits through queues,'' \emph{IEEE Transactions on
  Information Theory}, vol.~42, no.~1, pp. 4--18, 1996.

\bibitem{das_2000}
A.~Das, ``Capacity-achieving distributions for non-{Gaussian} additive noise
  channels,'' in \emph{2000 IEEE International Symposium on Information Theory
  (Cat. No.00CH37060)}, 2000, p. 432.

\bibitem{tchamkerten_discreteness_2004}
\BIBentryALTinterwordspacing
A.~Tchamkerten, ``\BIBforeignlanguage{en}{On the {Discreteness} of
  {Capacity}-{Achieving} {Distributions}},'' \emph{\BIBforeignlanguage{en}{IEEE
  Transactions on Information Theory}}, vol.~50, no.~11, pp. 2773--2778, Nov.
  2004. [Online]. Available: \url{http://ieeexplore.ieee.org/document/1347363/}
\BIBentrySTDinterwordspacing

\bibitem{Boyd_convex_optimization_2004}
S.~Boyd and L.~Vandenberghe, \emph{Convex Optimization}.\hskip 1em plus 0.5em
  minus 0.4em\relax Cambridge University Press, 2004.

\bibitem{Karush39}
W.~Karush, ``Minima of functions of several variables with inequalities as side
  conditions,'' Master's thesis, Department of Mathematics, University of
  Chicago, Chicago, IL, USA, 1939.

\bibitem{Kuhn51_481}
H.~W. Kuhn and A.~W. Tucker, ``Nonlinear programming,'' in \emph{Proceedings of
  the {S}econd {B}erkeley {S}ymposium on {M}athematical {S}tatistics and
  {P}robability, 1950}.\hskip 1em plus 0.5em minus 0.4em\relax Berkeley and Los
  Angeles: University of California Press, 1951, pp. 481--492.

\bibitem{blahut_1972}
R.~Blahut, ``\BIBforeignlanguage{en}{Computation of channel capacity and
  rate-distortion functions},'' \emph{\BIBforeignlanguage{en}{IEEE Transactions
  on Information Theory}}, vol.~18, no.~4, p. 460–473, Jul. 1972.

\bibitem{arimoto_1972}
S.~Arimoto, ``An algorithm for computing the capacity of arbitrary discrete
  memoryless channels,'' \emph{IEEE Transactions on Information Theory},
  vol.~18, no.~1, pp. 14--20, 1972.

\bibitem{dauwels_2006}
J.~H. Dauwels, ``\BIBforeignlanguage{en}{On graphical models for communications
  and machine learning: algorithms, bounds, and analog implementation},''
  Doctoral Thesis, ETH Zurich, Konstanz, 2006, diss., Eidgenössische
  Technische Hochschule ETH Zürich, Nr. 16365, 2005.

\bibitem{Witteveen25_arxiv}
\BIBentryALTinterwordspacing
O.~Witteveen, S.~J. Rosen, R.~S. Lach, M.~Z. Wilson, and M.~Bauer, ``Optimizing
  information transmission in the canonical wnt pathway,'' 2025. [Online].
  Available: \url{https://arxiv.org/abs/2506.22633}
\BIBentrySTDinterwordspacing

\bibitem{Rieke97}
F.~Rieke, D.~Warland, R.~{de Ruyter van Steveninck}, and W.~Bialek,
  \emph{Spikes: Exploring the Neural Code}.\hskip 1em plus 0.5em minus
  0.4em\relax Cambridge, MA: MIT Press, 1997.

\bibitem{Mattingly18_1760}
\BIBentryALTinterwordspacing
H.~H. Mattingly, M.~K. Transtrum, M.~C. Abbott, and B.~B. Machta, ``Maximizing
  the information learned from finite data selects a simple model,''
  \emph{Proceedings of the National Academy of Sciences}, vol. 115, no.~8, pp.
  1760--1765, 2018. [Online]. Available:
  \url{https://www.pnas.org/doi/abs/10.1073/pnas.1715306115}
\BIBentrySTDinterwordspacing

\bibitem{Fix78_704}
S.~L. Fix, ``Rate distortion functions for squared error distortion measures.''
  in \emph{Proc 16th Annu Allerton Conf Commun Control Comput}, pp. 704 -- 711.

\bibitem{Rose94_1939}
K.~Rose, ``A mapping approach to rate-distortion computation and analysis,''
  \emph{IEEE Transactions on Information Theory}, vol.~40, no.~6, pp.
  1939--1952, 1994.

\end{thebibliography}


%\begin{IEEEbiography}{Jonas Stapmanns}
%Likes pancakes.
%\end{IEEEbiography}

\end{document}